â\input harvmac
\noblackbox

\input epsf
\input psfig
\def\figin{\epsfcheck\figin}\def\figins{\epsfcheck\figins}
\def\epsfcheck{\ifx\epsfbox\UnDeFiNeD
\message{(NO epsf.tex, FIGURES WILL BE IGNORED)}
\gdef\figin##1{\vskip2in}\gdef\figins##1{\hskip.5in}
\else\message{(FIGURES WILL BE INCLUDED)}%
\gdef\figin##1{##1}\gdef\figins##1{##1}\fi}
\def\DefWarn#1{}
\def\figinsert{\goodbreak\midinsert}
\def\ifig#1#2#3{\DefWarn#1\xdef#1{fig.~\the\figno}
\writedef{#1\leftbracket fig.\noexpand~\the\figno}%
\figinsert\figin{\centerline{#3}}\medskip\centerline{\vbox{\baselineskip12pt
\advance\hsize by -1truein\noindent\footnotefont{\bf Fig.~\the\figno } #2}}
\bigskip\endinsert\global\advance\figno by1}

\def\tilde{\widetilde}


 \font\cmss=cmss10
 \font\cmsss=cmss10 at 7pt
 \def\rlx{\relax\leavevmode}
 \def\inbar{\vrule height1.5ex width.4pt depth0pt}
 \def\IC{\relax\,\hbox{$\inbar\kern-.3em{\rm C}$}}
 \def\IN{\relax{\rm I\kern-.18em N}}
 \def\IP{\relax{\rm I\kern-.18em P}}
 \def\ZZ{\rlx\leavevmode\ifmmode\mathchoice{\hbox{\cmss Z\kern-.4em Z}}
  {\hbox{\cmss Z\kern-.4em Z}}{\lower.9pt\hbox{\cmsss Z\kern-.36em Z}}
  {\lower1.2pt\hbox{\cmsss Z\kern-.36em Z}}\else{\cmss Z\kern-.4em
  Z}\fi}
 \def\IZ{\relax\ifmmode\mathchoice
 {\hbox{\cmss Z\kern-.4em Z}}{\hbox{\cmss Z\kern-.4em Z}}
 {\lower.9pt\hbox{\cmsss Z\kern-.4em Z}}
 {\lower1.2pt\hbox{\cmsss Z\kern-.4em Z}}\else{\cmss Z\kern-.4em
 Z}\fi}
 \def\IZ{\relax\ifmmode\mathchoice
 {\hbox{\cmss Z\kern-.4em Z}}{\hbox{\cmss Z\kern-.4em Z}}
 {\lower.9pt\hbox{\cmsss Z\kern-.4em Z}}
 {\lower1.2pt\hbox{\cmsss Z\kern-.4em Z}}\else{\cmss Z\kern-.4em Z}\fi}

 \def\narrowplus{\kern -.04truein + \kern -.03truein}
 \def\narrowminus{- \kern -.04truein}
 \def\narrowminussub{\kern -.02truein - \kern -.01truein}

 \def\frac#1#2{{#1\over #2}}

 \def\sym#1{{{\rm SYM}} _{#1 +1}}
 
 \def\IZ{\relax\ifmmode\mathchoice
 {\hbox{\cmss Z\kern-.4em Z}}{\hbox{\cmss Z\kern-.4em Z}}
 {\lower.9pt\hbox{\cmsss Z\kern-.4em Z}}
 {\lower1.2pt\hbox{\cmsss Z\kern-.4em Z}}\else{\cmss Z\kern-.4em Z}\fi}
 \def\IB{\relax{\rm I\kern-.18em B}}
 \def\IC{{\relax\hbox{$\inbar\kern-.3em{\rm C}$}}}
 \def\ID{\relax{\rm I\kern-.18em D}}
 \def\IE{\relax{\rm I\kern-.18em E}}
 \def\IF{\relax{\rm I\kern-.18em F}}
 \def\IG{\relax\hbox{$\inbar\kern-.3em{\rm G}$}}
 \def\IGa{\relax\hbox{${\rm I}\kern-.18em\Gamma$}}
 \def\IH{\relax{\rm I\kern-.18em H}}
 \def\II{\relax{\rm I\kern-.18em I}}
 \def\IK{\relax{\rm I\kern-.18em K}}
 \def\IP{\relax{\rm I\kern-.18em P}}

 \font\cmss=cmss10 \font\cmsss=cmss10 at 7pt
 \def\IR{\relax{\rm I\kern-.18em R}}

 %

 %
 %
 \def\eqnn#1{\xdef #1{(\secsym\the\meqno)}\writedef{#1\leftbracket#1}%
 \global\advance\meqno by1\wrlabeL#1}
 \def\eqna#1{\xdef #1##1{\hbox{$(\secsym\the\meqno##1)$}}
 \writedef{#1\numbersign1\leftbracket#1{\numbersign1}}%
 \global\advance\meqno by1\wrlabeL{#1$\{\}$}}
 \def\eqn#1#2{\xdef #1{(\secsym\the\meqno)}\writedef{#1\leftbracket#1}%
 \global\advance\meqno by1$$#2\eqno#1\eqlabeL#1$$}

\newdimen\tableauside\tableauside=1.0ex
\newdimen\tableaurule\tableaurule=0.4pt
\newdimen\tableaustep
\def\phantomhrule#1{\hbox{\vbox to0pt{\hrule height\tableaurule width#1\vss}}}
\def\phantomvrule#1{\vbox{\hbox to0pt{\vrule width\tableaurule height#1\hss}}}
\def\sqr{\vbox{%
  \phantomhrule\tableaustep
  \hbox{\phantomvrule\tableaustep\kern\tableaustep\phantomvrule\tableaustep}%
  \hbox{\vbox{\phantomhrule\tableauside}\kern-\tableaurule}}}
\def\squares#1{\hbox{\count0=#1\noindent\loop\sqr
  \advance\count0 by-1 \ifnum\count0>0\repeat}}
\def\tableau#1{\vcenter{\offinterlineskip
  \tableaustep=\tableauside\advance\tableaustep by-\tableaurule
  \kern\normallineskip\hbox
    {\kern\normallineskip\vbox
      {\gettableau#1 0 }%
     \kern\normallineskip\kern\tableaurule}%
  \kern\normallineskip\kern\tableaurule}}
\def\gettableau#1 {\ifnum#1=0\let\next=\null\else
  \squares{#1}\let\next=\gettableau\fi\next}

\tableauside=1.0ex
\tableaurule=0.4pt


\lref\amv{M.~Aganagic, M.~Mari\~no and C.~Vafa, ``All loop topological string
amplitudes from Chern-Simons theory,'' hep-th/0206164.}

\lref\AV{M. Aganagic and C. Vafa, ``G(2) manifolds, mirror symmetry 
and geometric engineering,'' hep-th/0110171.}
\lref\mv{M.~Aganagic and C.~Vafa, 
``Perturbative derivation of mirror symmetry,'' hep-th/0209138.}

\lref\silver{O.~Aharony, M.~Berkooz and E.~Silverstein, 
``Multiple-trace operators and non-local string theories,'' hep-th/0105309,
JHEP {\bf 0108} (2001) 006.}

\lref\HosonoXJ{
S.~Hosono,
``Counting BPS states via holomorphic anomaly equations,''
arXiv:hep-th/0206206.
}

\lref\CandelasDM{
P.~Candelas, X.~De La Ossa, A.~Font, S.~Katz and D.~R.~Morrison,
``Mirror symmetry for two parameter models. I,''
Nucl.\ Phys.\ B {\bf 416}, 481 (1994)
[arXiv:hep-th/9308083].
}

\lref\normal{
L.~L.~Chau and O.~Zaboronsky,
``On the structure of correlation functions in the normal matrix model,''
Commun.\ Math.\ Phys.\  {\bf 196}, 203 (1998)
[arXiv:hep-th/9711091].
}

\lref\arnoldbookII{V.I. Arnold, S.M. Gusein-Zade and A.N. Varchenko, 
{\sl Singularities of Differentiable Maps: Vol II}, Birh\"auser, Boston (1988).}

\lref\bcovII{
M.~Bershadsky, S.~Cecotti, H.~Ooguri and C.~Vafa,
``Kodaira-Spencer theory of gravity and exact results for quantum string
amplitudes,'' hep-th/9309140,
Commun.\ Math.\ Phys.\  {\bf 165}(1994) 311.} 

\lref\bcovI{
M.~Bershadsky, S.~Cecotti, H.~Ooguri and C.~Vafa,
``Holomorphic anomalies in topological field theories,'' hep-th/9302103, 
Nucl.\ Phys.\  {\bf B 405} (1993) 279.}

\lref\biz{D.~Bessis, C.~Itzykson and J.~B.~Zuber,
``Quantum field theory techniques in graphical enumeration,''
Adv.\ Appl.\ Math.\  {\bf 1} (1980) 109.}
\lref\AKV{M. Aganagic,
A. Klemm and C. Vafa, ``Disk instantons, mirror symmetry and the duality
web,'' hep-th/0105045, Z.\ Naturforsch.\ A {\bf 57}, 1 (2002)}
\lref\bd{G.~Bonnet, F.~David and B.~Eynard, 
``Breakdown of universality in multi-cut matrix models,'' cond-mat/0003324, 
J.\ Phys.\ {\bf A 33} (2000) 6739.} 

\lref\bipz{E.~Br\'ezin, C.~Itzykson, G.~Parisi and J.~B.~Zuber, 
``Planar diagrams,''
Commun.\ Math.\ Phys.\  {\bf 59} (1978) 35.}

\lref\ckyz{
T.~M.~Chiang, A.~Klemm, S.~T.~Yau and E.~Zaslow,
``Local mirror symmetry: Calculations and interpretations,'' hep-th/9903053,
Adv.\ Theor.\ Math.\ Phys.\  {\bf 3} (1999) 495.} 

\lref\dfg{D.~E.~Diaconescu, B.~Florea and A.~Grassi, ``Geometric
transitions and open string instantons,'' hep-th/0205234;
``Geometric transitions, del Pezzo surfaces and open string instantons,'' 
hep-th/0206163.}

\lref\difr{P. Di Francesco, 
``Matrix model combinatorics: applications to folding 
and coloring," math-ph/9911002.}

\lref\dfi{P.~Di Francesco and C.~Itzykson, 
``A generating function for fatgraphs,'' hep-th/9212108, 
Annales Poincare Phys.\ Theor.\  {\bf 59} (1993) 117.}

\lref\dva{R. Dijkgraaf and C. Vafa, ``Matrix models, 
topological strings, and supersymmetric gauge theories,'' hep-th/0206255,
Nucl.\ Phys.\ {\bf B 644} (2002) 3.
}

\lref\dvtwo{R.~Dijkgraaf and C.~Vafa, 
``On geometry and matrix models,'' hep-th/0207106,
Nucl.\ Phys.\ {\bf B 644} (2002) 21.}

\lref\dvthree{R.~Dijkgraaf and C.~Vafa, 
``A perturbative window into non-perturbative physics,'' hep-th/0208048.}

\lref\AVS{M. Aganagic and C. Vafa, ``Mirror symmetry, D-branes and
counting holomorphic discs,'' hep-th/0012041.}

\lref\dgkv{R.~Dijkgraaf, S.~Gukov, V.~A.~Kazakov and C.~Vafa, 
``Perturbative analysis of gauged matrix models,'' hep-th/0210238.}

\lref\ms{S.~Elitzur, G.~W.~Moore, A.~Schwimmer and N.~Seiberg, 
``Remarks on the canonical quantization of the Chern-Simons-Witten theory,''
Nucl.\ Phys.\  {\bf B 326} (1989) 108.}

\lref\gvtg{R. Gopakumar and C. Vafa, ``Topological gravity as
large $N$ topological
gauge theory," hep-th/9802016, Adv. Theor. Math. Phys. {\bf 2} (1998) 413.}

\lref\gvmone{R. Gopakumar and C. Vafa, ``M-theory and topological strings,
I," hep-th/9809187.}

\lref\gv{R. Gopakumar and C. Vafa, ``On the gauge theory/geometry
correspondence," hep-th/9811131, Adv. Theor. Math. Phys. {\bf 3} (1999)
1415.}

\lref\hata{S.K. Hansen and T. Takata, 
``Reshetikhin-Turaev invariants of Seifert
3-manifolds for classical simple Lie algebras,
and their asymptotic expansions,'' math.GT/0209043.}

\lref\hv{K. Hori and C. Vafa, ``Mirror symmetry,'' hep-th/0002222.}

\lref\iz{
C. Itzykson and J.~B.~Zuber, ``The planar approximation. 2,''
J.\ Math.\ Phys.\  {\bf 21} (1980) 411.}

\lref\kkvI{
S.~Katz, A.~Klemm and C.~Vafa,
``Geometric engineering of quantum field theories,'' hep-th/9609239, 
Nucl.\ Phys.\ {\bf B 497} (1997) 173.}

\lref\kz{
A.~Klemm and E.~Zaslow,
``Local mirror symmetry at higher genus,'' hep-th/9906046, in {\it Winter
School on Mirror Symmetry, Vector bundles and Lagrangian
Submanifolds}, p. 183, American Mathematical Society 2001.}

\lref\kon{M. Kontsevich, ``Intersection theory on the moduli space of
curves and the matrix Airy function," Commun. Math. Phys. {\bf 147} (1992)
1.}

\lref\km{I.K. Kostov and M.L. Mehta, ``Random surfaces of arbitrary genus:
exact results for $D=0$ and $-2$ dimensions," Phys. Lett. {\bf B 189}
(1987) 118.}

\lref\morecan{J.M.F. Labastida and A.~V.~Ramallo, 
``Operator formalism for Chern-Simons Theories,''
Phys.\ Lett.\ {\bf B 227} (1989) 92.
J.M.F.~Labastida, P.~M.~Llatas and A.~V.~Ramallo,
``Knot operators in Chern-Simons gauge theory,''
Nucl.\ Phys.\ {\bf B 348} (1991) 651.}

\lref\mm{M. Mari\~no, ``Chern-Simons theory, matrix integrals, and
perturbative three-manifold invariants,'' hep-th/0207096.}
\lref\ovknot{H. Ooguri and C. Vafa, ``Knot invariants and topological
strings," hep-th/9912123, Nucl. Phys. {\bf B 577} (2000) 419.}

\lref\ov{H. Ooguri and C. Vafa, ``Worldsheet derivation of a large $N$
duality,'' hep-th/0205297, Nucl. Phys. {\bf B 641} (2002) 3.}

\lref\roz{L. Rozansky, ``A large $k$ asymptotics
of Witten's invariant of Seifert
manifolds,'' hep-th/9303099,
Commun.\ Math.\ Phys.\  {\bf 171} (1995) 279.
``A contribution of the trivial connection to Jones
polynomial and Witten's invariant of 3-D manifolds. 1'' hep-th/9401061,
Commun.\ Math.\ Phys.\  {\bf 175} (1996) 275.}

\lref\sts{C. Vafa, ``Superstrings and topological strings at large $N$,''
hep-th/0008142, J.\ Math.\ Phys.\  {\bf 42} (2001) 2798.}

\lref\jones{E. Witten, ``Quantum field theory and the Jones polynomial,"
Commun. Math. Phys. {\bf 121} (1989) 351.}
\lref\wittcs{E. Witten, ``Chern-Simons gauge theory as
a string theory,'' hep-th/9207094, in {\it The Floer memorial volume},
H. Hofer, C.H. Taubes, A. Weinstein and E. Zehner, eds.,
Birkh\"auser 1995, p. 637.}
\lref\multi{E.~Witten, 
``Multi-trace operators, boundary conditions, and AdS/CFT correspondence,''
hep-th/0112258.}
\lref\hp{A. Giveon, A. Kehagias and H. Partouche, ``Geometric transitions,
brane dynamics and gauge theories,'' hep-th/0110115, 
JHEP {\bf 0112} (2001) 021.}
%
\lref\wb{
E.~Witten,
``Branes and the dynamics of QCD,''
Nucl.\ Phys.\ B {\bf 507}, 658 (1997)
[arXiv:hep-th/9706109].
}


\Title
 {\vbox{
 \baselineskip12pt
 \hbox{hep-th/0211098}\hbox{HUTP-02/A057}\hbox{HU-EP-02/47}
}}
 {\vbox{
 \centerline{Matrix Model as a Mirror of Chern-Simons Theory}
 }}

 \centerline{ Mina Aganagic,$^{a}$ 
Albrecht Klemm,$^{b}$ Marcos Mari\~no,$^{a}$ and 
Cumrun Vafa$^{a}$}
 \bigskip
\centerline{$^a$ Jefferson Physical Laboratory, Harvard University}
\centerline{Cambridge, MA 02138, USA}

\centerline{$^b$ Humboldt-Universit\"at zu Berlin, Institut f\"ur
Physik}
\centerline{D-10115 Berlin, Germany}
 \smallskip
 \vskip .3in \centerline{\bf Abstract}
\vskip .2in
 { Using mirror symmetry, we show that
Chern-Simons theory on certain manifolds such as lens spaces reduces to a novel
class of Hermitian matrix models, where the measure is that of unitary
matrix models.  We show that this agrees with the more conventional
canonical quantization of Chern-Simons theory. Moreover, large $N$ dualities
in this context lead to computation of all genus A-model topological amplitudes
on toric Calabi-Yau manifolds in terms of matrix integrals. In the context
of type IIA
superstring compactifications on these Calabi-Yau manifolds with wrapped D6 branes (which are dual to M-theory on $G_2$ manifolds)
this leads to engineering and solving F-terms for ${\cal N}=1$
supersymmetric gauge theories with superpotentials involving certain
multi-trace operators.
}
 \smallskip \Date{}

\newsec{Introduction}
Recently it was observed in \mm\ that partition functions of Chern-Simons
theory on certain manifolds can be represented as Hermitian matrix integrals
with a measure suitable for unitary matrix models.  On the other hand,
it was found in \dva\ that topological strings for B-branes are equivalent to
Hermitian matrix models.  It is thus natural to ask if these two
ideas are related.  Since Chern-Simons theory arises
from topological strings for A-branes \wittcs\ one is led to believe
that the observation in \mm\ should be obtained by applying mirror symmetry
to obtain certain B-brane matrix models. In this paper we will verify that this
is indeed the case. For example by applying mirror symmetry
to the deformed conifold $T^* {\bf S}^3$ 
we show that the Chern-Simons theory on ${\bf S}^3$
reduces to a Gaussian Hermitian matrix model with a unitary measure.

On the other hand the large $N$ transition
proposed in \gv , and derived from the worldsheet viewpoint in \ov ,
relates Chern-Simons gauge theory to A-model topological strings (with
or without branes) on certain
non-compact Calabi-Yau threefolds. Thus the result we obtain here 
shows that the
topological A-model on certain non-compact Calabi-Yau manifolds
reduces to matrix integrals.
In particular we consider the $\IZ_p$ orbifold of the duality
in \gv\ which suggests that Chern-Simons theory on lens space should
be related to the $\IZ_p$ quotient of the resolved conifold\foot{This
idea has been advanced by a number of physicists, including
R. Gopakumar, S. Sinha, E. Diaconescu, A. Grassi, B. Pioline, J. Gomis
and E. Cheung. See also \hp.}. We find that the large
$N$ duality continues to hold upon orbifolding, and the
choice of flat connection in the Chern-Simons theory on lens
space maps to the extra blowup moduli from the twisted sectors 
on the closed string side.

This is a natural extension
of the result that matrix integrals can compute intersection theory
on moduli space of Riemann surfaces \kon.  Moreover
this sheds a new light on recent results \refs{\amv ,\dfg} which
relate all genus open and closed topological A-model
amplitudes with Chern-Simons
theory.  Namely, we can restate (and rederive) this result
in terms of the equivalence of topological A-model and a suitable
matrix model.

The matrix model we end up with is a novel kind of matrix model, 
in which the action is that of a Hermitian matrix model $V(u)$, but
the measure is that suitable for a unitary matrix $U=e^u$.  This is not a
unitary matrix model. In particular
the action does not have the periodicity expected for a unitary matrix model.
We explain how this arises from mirror symmetry.  Moreover we are able
to rewrite this in terms of an ordinary Hermitian matrix model
with the usual measure, at the expense of introducing multi-trace operators
in the action.

From the viewpoint of type IIA compactifications the A-branes which
fill spacetime give rise to ${\cal N}=1$ supersymmetric gauge theories.
For example $N$ D6 branes wrapped on ${\bf S}^3\subset T^*{\bf S}^3$ gives 
rise,
in the infrared, to pure $U(N)$ Yang-Mills theory.  However the F-terms
of the full theory differ from that of pure Yang-Mills.  
Here we find, using this
rewriting of the measure, that the theory can be viewed as a deformed
${\cal N}=2$ theory with a mass term for the adjoint $m\, {\rm Tr}\, 
 \Phi^2$, together
with certain multi-trace operators of the form $S \, {\rm Tr}\,  \Phi^k \,  
{\rm Tr}\, \Phi^l$
where $S$ is the glueball field $S={\rm Tr} {\cal W}^2$. Thus we can capture
the deviations from the pure Yang-Mills in terms of these multi-trace
operators.  Note that, upon lifting to M-theory, these theories give an
effective description of ${\cal N}=1$ compactifications of
M-theory on certain $G_2$ manifolds.
\ifig\diagram{Interrelations of various topics covered in this paper.}
{\epsfxsize5.0truein\epsfbox{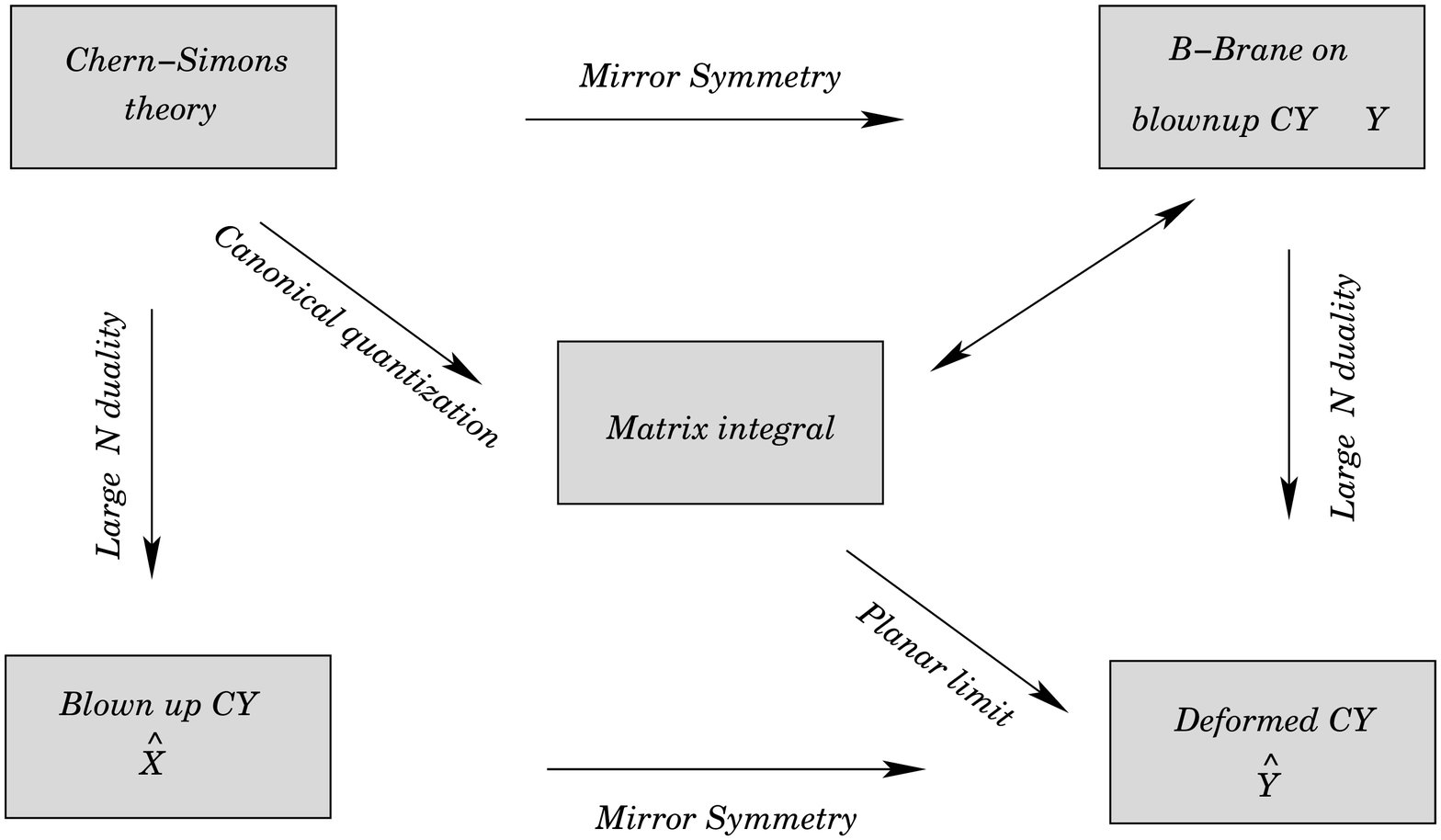}}
The organization of this paper is as follows: in section 2, we review
Chern-Simons theory and how it arises 
in the context of A-model topological strings. In particular 
we show that the matrix model expression of \mm\ for the
partition function is already natural from the point of view of canonical
quantization of Chern-Simons theory. 
In section 3, we present the mirror to the A-model 
geometries, following the ideas in \AV. We
also analyze the topological theory describing B-branes in the mirror
geometry in the spirit of \dva, and we show that it reduces to a matrix
model. This provides a mirror symmetry derivation of the Chern-Simons
matrix models advanced in \mm. In section 4, we show that the standard
planar limit analysis \bipz\ of the matrix model leads to the mirror of the 
deformed conifold, showing in this way that the large $N$ limit of the
Chern-Simons matrix model leads naturally to the mirror of the 
large $N$ transition
proposed in \gv. In section 5, we extend the analysis to the case of lens
spaces. We again give a mirror symmetry derivation of the corresponding
matrix model describing Chern-Simons theory, and give a detailed
comparison with standard results in Chern-Simons theory. Furthermore, we extend
the large $N$ duality to the orbifolds of \gv\ by $\IZ_p$.
We do a detailed perturbative computation for $p=2$, by rewriting
the Chern-Simons matrix model for lens spaces 
as a Hermitian matrix model.
In section 6, we consider the closed string geometry which is the large $N$
dual of $T^*( {\bf S}^3/\IZ_2)$, namely local $\IP^1 \times \IP^1$. We give a
fairly complete description of the extended K\"ahler moduli space and we
compute the $F_g$ couplings by using the B-model Kodaira-Spencer theory of
\bcovII. In order to test the large $N$ duality, we expand these coupling
around the point in moduli space where both $\IP^1$'s have
vanishing quantum volume, and find perfect agreement with the results of matrix
model/Chern-Simons perturbation theory. In section 7, we present some
generalizations of the mirror symmetry derivation of the matrix model. In
particular, we show how to include matter, making in this way contact with
the results of \refs{\amv,\dfg}. Finally, in section 8 we put our results in
the context of type IIA compactifications with spacetime filling branes,
and we show that the resulting gauge theories include multi-trace
operators that can be read off from the Hermitian matrix model of section 5. 
Finally, the two appendices collect some useful results on computation of 
averages in the Gaussian matrix model, and on the solution of
the holomorphic anomaly equation. 

\newsec{Physics of the A-model and Chern-Simons Theory}

As shown in \wittcs, if we wrap
$N$ D-branes on $M$ in $T^* M$, the associated topological A-model
is a $U(N)$ Chern Simons theory on the three-manifold $M$
\eqn\pathcs{Z= \int {\cal D} A e^{S_{\rm CS}(A)}}
where
$$S_{\rm CS}(A) = \frac{i k}{4 \pi}\int_{M} {\rm Tr} (A\wedge dA
+ \frac{2}{3} A\wedge A\wedge A)$$
is the Chern-Simons action. 
The basic
idea of this equivalence is as follows.
The path-integral of the topological A-model
localizes on holomorphic curves, and when there are D-branes, this
means holomorphic curves with boundaries
ending on them. In the $T^*M$ geometry
with D-branes wrapping $M$ there are no honest holomorphic curves,
however there are degenerate holomorphic curves that look like trivalent
ribbon graphs and come from the boundaries of the moduli space.
This leads to a field theory
description in target space, which is equivalent to topological
Chern-Simons theory.
In this map, the level $k$ would be naively related
to the inverse of the string coupling constant $g_s$.
However, quantum corrections shift this identification to
$$\frac{2\pi i}{k+N} = g_s.$$
The perturbative open-string expansion and Chern-Simons ribbon
graph expansion around their classical vacua coincide.

In this paper we mainly consider $M$'s that are $T^2$ fibered over an
interval $I$. The fiber over a generic point in $I$ is a $T^2$, but
some $(p,q)$ one-cycles of the $T^2$ degenerate at the end points. 
Alternatively, we can view $M$ as obtained by
gluing two solid tori $T_L$ and $T_R$ over the midpoint of
the interval, up to an ${\rm SL}(2,{\bf Z})$ transformation $U$
that corresponds to a diffeomorphism identification of their boundaries.
Let $(p_L,q_L)$ be the cycle of the $T^2$ fiber
that degenerates over the left half on $M$, and let $(p_R,q_R)$
be the cycle that degenerates over the right half.
The gluing matrix $U$ can be written as
\eqn\glue{
U = U_L^{-1} U_R,}
where $U_{L,R} = \left(\matrix{p_{L,R} & s_{L,R}\cr
q_{L,R} &t_{L,R}}\right)\in SL(2,{\bf Z})$.
(Clearly, $U$ is unique up to a homeomorphism that changes the ``framing''
of three-manifold \jones\ and takes
\eqn\fram{V_{L,R}\rightarrow V_{L,R}\; T^{n_{L,R}}}
where $T$ is a generator of $SL(2,\bf{Z})$,
$T=\left(\matrix{1&1\cr 0&1}\right)$.
This is a consequence of the fact that there is no natural choice of
the cycle that is finite on the solid torus. We will come back to this
later.)

Consider $M$ with an insertion of a
Wilson line in representation $R$ in $T_L$, and
a Wilson line in representation $R'$ in $T_R$ along
the one-cycles of the solid tori that are not filled in.
The partition function is given by
$$Z(M; R, R') = \langle R| U|R'\rangle.$$
Above, $| R\rangle$ for example, corresponds to computing the path integral on 
the solid torus $T_L$. Moreover, it 
gives a state in the Hilbert space of Chern-Simons 
theory on $T^2$ on the boundary of $T_L$.
The $SL(2,\IZ)$ transformations
of the boundary act as operators on this Hilbert space.
The corresponding states and operators
can be found by considering canonical 
quantization of Chern-Simons theory on $M=T^2\times R$, 
following \ms\ (see also \morecan). 
This allows one to solve the theory, and in particular
to show that the theory is equivalent to a matrix model.
Let us begin by briefly recalling \ms.

By integrating over $A_t$ where the time $t$ corresponds to the
$R$ direction in $T^2\times R$, the Chern-Simons
path integral becomes
\eqn\cst{ Z = \int{ \cal D}A_{u}{ \cal D}A_{v}\; \delta(F_{uv}) \; \exp \Bigl(\frac{k}{2
\pi i}\int_{M} {\rm Tr} A_v \dot{A}_u \Bigr).}
The delta function localizes to $A$'s which are flat connections on the $T^2$.
As the fundamental group of the 2-torus is commutative, 
by a gauge transformation, we can
set $A = u\; d \theta_u + v\; d \theta_v$ where
$u$ and $v$ are holonomies of the gauge field
along the $(1,0)$ and $(0,1)$ cycle of the $T^2$.
Integrating out the unphysical degrees of freedom is rather
subtle, but the main physical effect is
to incorporate the shift of $k\rightarrow \hat k = k+N$.
Thus, we can simply consider the naive quantization, with $k$
replaced by $\hat{k}$ -- the effective value of $k$ is also
what enters in the string coupling constant $g_s$.

We can now construct the operators representing
the action of $SL(2,\IZ)$ on the Hilbert space of $T^2$, by 
noting that $u$ and $v$ are conjugate variables,
with
$$[u_{i},v_{j}] = g_s \delta_{ij}.$$
The action of $S$ and $T$ operators on the $T^2$ implies that 
$$T : \ u \rightarrow u+v, \ \  v \rightarrow v\, ;\ \ \ \ \  S : \ v
\rightarrow u, \ \ u \rightarrow -v,$$
and this suffices to determine them up to normalization \ms:
\eqn\slt{T = \eta_T\; {\rm e}^{-{\rm Tr} \, v^2/2 g_s};\ \ \ \ \ S =\eta_S\;
{\rm e}^{-{\rm Tr}(u^2 + v^2)/ 4 \pi g_s}.}

Suppose that the $v$-cycle of the $T^2$ is the one that is filled in.
The wave function corresponding to the the path integral
on the solid torus with insertion of a Wilson line in representation $R$
along the cycle which is finite is given by
\eqn\wave{\langle v | R_v \rangle =\frac{1}{|{\cal W}|}\sum_{w \in {\cal W}
} \epsilon(w) \delta(v + i g_s\omega(\alpha_R)).}
The sum is over the elements $w$ of the Weyl group 
where $\epsilon (w)$ is their signature. For $U(N)$ the order of the Weyl group
is $|{\cal W}|= N!$. Moreover, $\alpha_R$ 
is the highest weight vector of representation $R$, shifted by
the Weyl vector $\rho=\frac{1}{2}\sum_{\alpha>0}\alpha$
with $\alpha>0$ corresponding to positive roots. 
In particular, for the partition function without
any insertions $\alpha_{0}=\rho$. 

In writing the wave function in equation \wave\
we do not divide by the full group of large gauge transformations
on the $T^2$, but only by the Weyl group\foot{In that the equation \wave\ 
differs from equation $4.12$ of \ms.}.
The path integral 
on the solid torus can be viewed as a path integral on an interval where
$v$ is frozen at the end-point where the $v$-circle is filled, and 
the large gauge transformations that shift $v$ by $2\pi \alpha$ for
$\alpha$ in the root lattice $\Lambda_R$ are not a symmetry. 
In fact, generically the large gauge transformations
are broken to the Weyl group by the operators in \slt . 
This will be more transparent yet in the mirror B-model
language.

%
%
Consider for example the partition function on a three-manifold
$M$ where $(p_L,q_L) = (0,1)$ and $(p_R,q_R) = (1,1)$, with no insertions.
The gluing operator is $U=TST$, takes
$v$ to $u+v$, and leaves $u$ invariant. In terms of $u$ and $v$ it is 
given by $U = \exp({\rm Tr}\, u^2/2 g_s)$, up to normalization.
Correspondingly, we have
\eqn\waveav{Z(M) = \langle 0_{v}|\exp({\rm Tr}\,  u^2/2 g_s)| 0_v\rangle,}
where $|0_v \rangle$ is the partition function on a solid
torus with no insertions.
By writing $|0_v\rangle$ in the $u$ basis,
we see that the theory can be described by a
matrix model in terms of $u$, ${\rm e}^{iu}\in U(N)$
\eqn\pathsa{Z = \frac{1}{{\rm vol}(U(N))}\int d_H {u} \, \, \exp({\rm Tr}
\, u^2/2 g_s))}
where $d_H u$ is the Haar measure on $U(N)$. To show this, note that 
$$\langle u| 0_v \rangle=\Delta_{H}(u)=\prod_{\alpha>0} 2 \sin\Bigl({\alpha \cdot u\over 2}\Bigr),$$ 
where we used Weyl denominator formula
$\sum_{w \in {\cal W}} \epsilon(w) \exp(w(\rho) \cdot u)=
\prod_{\alpha>0} 2 \sinh\Bigl({\alpha \cdot u\over 2}\Bigr).$  
Recall that the positive roots of $U(N)$ are given by $\alpha_{ij}=e_i
-e_j$, for $i<j$ where $e_j$ form an orthonormal basis, and 
$\alpha_{ij}\cdot u=u_i -u_j$.  
On the other hand, it is
a well known result that the Haar measure on $U(N)$ becomes, when 
expressed in terms of the eigenvalues, 
\eqn\eiga{\frac{1}{{\rm vol}(U(N))}
\int d_H {u} = \frac{1}{|{\cal W}|} \int \prod_i d u_i\; \Delta^2_H (u),}
upon integrating over angles. Therefore, \waveav\ equals \pathsa. Notice
that, since we are not dividing by large gauge transformations, the
integration region for the eigenvalues $u_i$ is $\IR^N$.  

We can evaluate \pathsa\ explicitly by 
using the Weyl denominator formula to
rewrite \eiga\ as a Gaussian integral. We find
\eqn\check{Z = (- 2 \pi g_s)^{N/2} \eta_U
\sum_{w \in \cal
W}\epsilon(w)e^{\frac{g_s}{2}
(\rho+w(\rho))^2.}}
In the equation above, we denoted by $\eta_U$ the normalization of the $U=TST$
operator which we have not fixed.

In \jones, Chern-Simons theory was solved by relating the
Hilbert space of Chern-Simons theory
to the space of conformal blocks of
WZW model. The action of ${\rm SL}(2,\IZ)$ on the
conformal blocks of WZW model allows one to read off the matrix
elements of the operator corresponding to $U$. We will now show that the above
matrix model formulation agrees with the known results for 
$U(N)_k$ WZW model on ${\bf S}^3$ with the corresponding framing.
Namely, consider $U$ corresponding to the
$SL(2,\IZ)$ matrix
\eqn\generalmat{
U =\pmatrix{a & r \cr
b & s}.}
The path integral with Wilson lines in representation labeled by
$\alpha_R$, $\alpha_{R'}$ inserted parallel to the axis of the 
solid tori before the gluing with $U$ are given by \roz\hata\
\eqn\matsl{
\langle R| U| R' \rangle  =c_U
\sum_{n \in \Lambda_{\rm r}/b \Lambda_{\rm r}}
\sum_{w \in {\cal W}} \epsilon (w) \exp
\Bigl\{ {i \pi \over \hat{k} b} ( a \alpha_{R}^2 - 2\alpha_{R} \cdot (\hat{k} n + w(\alpha_{R'}))
+ s(\hat{k} n + w(\alpha_{R'}))^2 )\Bigr\}.}
We recall that $\hat k =k+N$, and the coefficient $c_U$ is given by
\eqn\coeff{c_U= {[i \, {\rm sign}(b)]^{N(N-1)/2} \over (\hat{k} |b|)^{N/2}}
\exp \Bigl[ -{ i (N^2-1)\pi \over 12}  \Phi (U)\Bigr],}
that only depends on $U$ and not on the Wilson-lines. Above,
$\Phi (U)$ is the Rademacher function:
\eqn\rade{
\Phi\left[ \pmatrix{ a & r \cr b&s}  \right]=
{a + s \over b} - 12 s(a,b),}
where $s(a,b)$ is the Dedekind sum
$$
s(a,b)={1 \over 4b} \sum_{n=1}^{b-1} \cot \Bigl( {\pi n \over b}\Bigr)
\cot \Bigl( {\pi n a\over b}\Bigr).
$$
In particular, we see that the partition function on $S^3$
corresponding to $U=TST$ agrees
with the expression we found above, provided we identify
$$\eta_{TST} = \frac{1}{(2\pi)^N}e^{-\frac{2 \pi i (N^2-1)}{12}}.$$
We will make many further checks of this formalism in the following
sections (In particular we will check that arbitrary matrix 
elements of $U$ agree with \matsl.).

\newsec{Mirror symmetry}

\subsec{Mirror Pairs of Geometries}

As discussed above, Chern-Simons theory on a three-manifold
$M$ is the same as topological A-model string on $T^*M$.
When $M$ is a $T^2$ fibration over an interval, the geometry
of $X=T^*M$ is rather simple.
As shown in \AV, $X$ itself is a Lagrangian $T^2\times \IR$ 
fibration with base $\IR^3$,
and where one-cycles of the $T^2$ degenerate over lines in the base.
Moreover the $T^2$ fiber of $X$ and the fiber $M$ can be identified.
In the Calabi-Yau geometry, there is a natural choice of basis
of $(1,0),(0,1)$ cycles of the $T^2$
that fibers $X$, which is provided by the choice of complex structure on $X$.
We can identify the one-cycles of the $T^2$ fiber
that shrink over the left and the right sides of the interval with
the shrinking 1-cycles of $T_L$ and $T_R$.
The diffeomorphism map $U$ is the ${\rm SL}(2,{\bf Z})$ transformation
that relates one of the shrinking cycles
of the fiber of $X$ to the other one.
Moreover, while any path between the lines in $\IR^3$ lifts to a 
three-manifold in $X$,
the path of minimal length lifts to $M$.

For example, $X=T^* {\bf S}^3$ can be written as
\eqn\con{xu + yv = \mu.}
The $T^2$ fiber of $X$ is visible from the fact that
the equation is invariant under $U(1)^2$ action where
$x,u$ are charged oppositely under the first
and $y,v$ under the second $U(1)$.
The minimal ${\bf S}^3$ embeds via $u=\bar x$ and $v = \bar y$,
$$|x|^2+|y|^2 = \mu,$$
and if $\mu$ is real and positive this is a three-sphere.
In view of the discussion above, we can regard this ${\bf S}^3$ as a real interval,
together with the $(1,0)$ one-cycle of the $T^2$ fiber that corresponds to the
phase of $x$ and the $(0,1)$ cycle
that is the phase of $y$. Alternatively, we have the gluing operator $U=S$.
The $(1,0)$ and $(0,1)$ cycles degenerate
over the $x=0$ and $y=0$ endpoints of the interval, respectively,
and these are two copies of $\;\IC^*$ in $X$ -- holomorphic cylinders
$\IR \times {\bf S}^1$.

As shown in \AV, manifolds mirror to the above Calabi-Yau geometries
can be obtained by deformation of the mirror duality proven in
\refs{\hv,\mv}. We refer the reader to \refs{\AV,\amv} for the details of this and here simply 
state the result.
Suppose $M$, viewed as a $T^2$ fibration,
has $(p_L,q_L)$ and $(p_R,q_R)$ cycles of the $T^2$ which degenerate over
the boundaries of the base interval. Correspondingly, $X$ has two
lines of degenerate fibers in the base.
The mirror manifold of $X$, we will call it $Y$, is given by resolution of
the following singularity 
\eqn\mirr{xy = P_L(u,v) P_R(u,v),}
where
\eqn\mirrlens{P_L = e^{p_L u + q_L v} -1, \quad P_R =e^{p_R u + q_R v} -1.}
Above, $u$ and $v$ are $\;\IC^*$ valued, so their imaginary
parts are periodic, with period $2\pi$.
The resolution is by blowing up the locus $x=y=0 = P_L = P_R,$
by inserting a $\IP^1$. If $z,z'$ are coordinates on the $\IP^1$,
$z=1/z'$ the resolution corresponds to covering $X$ by two
patches $X_L$ and $X_R$ given respectively
by
$$ (L) \quad xz = P_L \quad, \quad \quad (R) \quad yz' = P_R,$$ 
in $x-z- u-v$ coordinates for $X_L$ and in $y-z'-u-v$ space for $X_R$.
The transition functions are obvious, relating e.g. $y = P_R z$.

The minimal holomorphic $\IP^1$ is where one is blowing up.
This can be deformed to an ${\bf S}^2$ that is generally not holomorphic
by letting $x,y,u,v$ be arbitrary functions of $z,\bar z$ coordinates on
the sphere, obeying above transition functions.
However, the allowed deformations are not entirely arbitrary,
as the equation of $Y$ restricts the north pole
of the ${\bf S}^2$ ($z=0$) and the south pole ($z'=0$) to lie at 
$$(L) \quad P_L =0 \quad,\quad\quad (R) \quad P_R=0.$$
These deformations mirror the deformations of $M$ in $X$.
Topologically, $M$ comes in a family of 3-submanifolds
of $X$, by deforming the path in the base connecting the two lines
arbitrarily, and the condition on the north and the south pole of
the $M$ to lie on the lines in base of $X$ replaced by the above
holomorphic constraint on the mirror two-spheres.
This is natural in the view of the fact (which one can show using \hv) that 
the imaginary parts of $u$ and $v$ in the B-model
are T-dual to the 1-cycles of the $T^2$ in the A-model \refs{\AV,\amv}.

For example, the mirror of $T^*{\bf S}^3$ in \con\ is given by
blowup of
$$xy = (e^{u} - 1)(e^{v} -1)$$
as described above.
Mirror symmetry relates $N$ D-branes wrapping the ${\bf S}^3$ in the
A-model to $N$ B-branes
wrapping the $\IP^1$ in the mirror geometry
\foot{The subtlety regarding the choice of
framing of the three-manifold in $X$
is related in part to performing global ${\rm SL}(2,\IZ)$ transformations
of the $T^2$ fiber, which is a symmetry of the A-model theory.
There is a similar subtlety in defining the
B-model \AKV,
and part of the framing ambiguity that can be traded for an ${\rm SL}(2,Z)$
transformation of the geometry corresponds in the $B$ model
to transformation that takes $Y$ to
$xy = (e^{u+ mv } - 1)(e^{v} -1).$
}.

\subsec{The mirror B-model D-branes}

In this section we consider B-branes,
wrapping $\IP^1$'s in the B-model geometries described above.
We will show that the B-model theory is
described by a matrix model, as in \dva, albeit of a novel kind.
By mirror symmetry, the B-branes on $Y$ and the A-branes on $X$
should give rise to the same theory. We will show that 
the matrix model describing the B-branes at hand is precisely the 
matrix model we arrived upon in section 2, by considering canonical
quantization of Chern-Simons theory, and consequently
the same matrix model as in \mm.

In the simplest example, with $(p_{L,R},q_{L,R} ) =(0,1)$ the
manifold $Y$ is given by blowing up
$$ xy = (e^{v}-1)^2, \quad \quad u.$$
This contains a family of $\IP^1$'s parameterized by $u$,
and is mirror to A-model geometry containing a family of
${\bf S}^2\times {\bf S}^1$'s.
Above, $u$ and $v$ are $\;\IC^*$ valued, so their imaginary
parts are periodic, with period $2\pi$ \foot{Note that if we did forget about compactness of $v$ and of $u$
the above geometry would be an $A_1$ ALE space times $\;\IC$.}.

We can choose to parameterize the normal directions
to D-branes by $v$ and $u$, and in terms of these, the action on the
$N$ D-branes wrapped on a $\IP^1$ in this geometry is given by
\eqn\act{S =  \int_{\IP^1} {\rm Tr}{ {v}^{(1)} \bar{D}{u} },}
where ${v}^{(1)} = {v} /z dz$ is a one-form on $\IP^1$ valued in
the
Lie algebra of $U(N)$, and $\bar{D} = \bar{\partial} + [A,\;\;]$ for
$A$ a holomorphic $U(N)$ connection on the $\IP^1$.
Note that $v$ is a section of the trivial bundle on the $\IP^1$
as $e^v = xz+1$ is globally defined on $Y$, and the same is true for $u$.
The action is a non-Abelian generalization
of
$$S  = \int_{B(\cal{C,C}_*)} \Omega$$
the action for a single D-brane on the $\IP^1$ \dva. 
Above, $\Omega = \frac{dv dz du}{z}$
is the holomorphic three-form on $Y$\foot{Here, $v(z,\bar z)$ and $u(z,\bar z)$ are viewed as maps deforming
the holomorphic curve ${\cal C}_*$ to a nearby curve ${\cal C}$ which is not
holomorphic, and $B({\cal C,C}_*)$ is the 3-chain interpolating between
them. Evaluated for an infinitesimal deformation along
the $v$ direction, this gives the action \act\ for a single D-brane.}.
As a further check, note that the 
equations of motion corresponding to the action \act\ have solutions
which agree with the geometric picture. That is
$${\bar D}{ u} = 0 = {\bar D} ({v}/z dz),$$
is solved by $u$ an arbitrary constant on the D-brane, and moreover
the $v$ equation of motion requires
$v \sim z$ near the north pole $z=0$ and $v \sim z'$ near the south pole
$z'=0$, and is therefore zero throughout.
In terms of the path-integral,
the action localizes on the paths for which
$v$ vanishes on the north and the south poles
of the sphere, and the equations of the blowup imply this as well.

Note that \act\ is the same as the action of Chern-Simons in the temporal
gauge, provided we identify the holonomies around the two
1-cycles of the $T^2$ in Chern-Simons.
In fact mirror symmetry
provides this identification naturally! Since the $T^2$ 
in the B-model, corresponding to the imaginary parts $u,v$
variables in $Y$ being compact, is mirror to the $T^2$
that fibers $X$, the identification of variables 
above follows simply by applying
T-duality on the D-branes (To be precise, in comparing
to \cst\ one should also
replace the $\IP^1$ by a cylinder, by replacing $dz/z = d \rho$, where
the cylinder is parameterized by $\rho$.).

For more general three-manifolds \eqns{\mirr,\mirrlens},
%
%
the north and the south pole of the D-brane are
constrained to live on $p_Lu + q_L v=0$ and $p_R u+q_R v =0$.
We can think of the theory on the D-brane 
as obtained 
by gluing together two
halves of $\IP^1$'s \dva . The action on both halves is the 
same, as the holomorphic three-form $\Omega$ is the same,
but there is a non-trivial map between the two boundaries.
That is, writing the partition function on the $\IP^1$ as 
$Z = \langle \rho_{L}| \rho_{R} \rangle,$
the states $|\rho_{L,R}\rangle$ are obtained by evaluating
the path integral over the north and the south cap of the $\IP^1$.
In the present context, these correspond to imposing the boundary conditions 
$p_{L,R}u+q_{L,R} v=0$, classically, so we can denote
$$|\rho_{L,R} \rangle = |\, 0_{p_{L,R} u+q_{L,R} v}\rangle,$$
therefore
%
$$Z = \langle 0_{p_L u+q_L v}\,|\, 0_{p_R u+q_Rv}\rangle.$$
Note that $u$ and $v$ are conjugate variables in the Lagrangian, 
so if we know $|0_v\rangle$, the state corresponding
to $|\, 0_{p u+ q v}\rangle$ is related to it by an operator $U$ 
$$| 0_{pu+qv} \rangle = U_{(p,q)}\, | 0_{v}\rangle,$$
such that 
$$U_{(p,q)} v U_{(p,q)}^{-1} = pu+qv,$$
as discussed above in the Chern-Simons context.

Moreover, in the present context, the effective gluing operator,
$U =  U_{(p_L,q_L)}^{-1} U_{(p_R,q_R)},$
should be naturally related to the superpotential $W$ of the theory.
Namely, the 
operator $U$ encodes difference of boundary conditions on the north and
the south poles of the $\IP^1$ which is what makes the supersymmetric vacua
in the generic
geometry \mirr\ isolated. In turn, this is precisely what the
superpotential $W$ encodes.
As an example, consider $v=0$ as the boundary conditions
on the left half of the $\IP^1$, and 
$u+v=0$ on the right,
corresponding to a B-brane on 
\eqn\mirrb{ xz = (e^{v} -1 ) ( e^{v+u } -1).}
Then, 
$U =  \exp(\frac{1}{2g_s} \int_{\IP^1} \omega\, {\rm Tr}\, u^2  ),$
where $\omega$ is a $(1,1)$ form on $\IP^1$ of unit volume.  
The blowup of the manifold in 
\mirrb\ corresponds to mirror of $T^*{{\bf S}^3}$ with non-trivial framing 
that we studied in detail in section 2. 
In fact, 
$$U =  \exp\Bigl(\frac{1}{g_s} \int_{\IP^1} \omega\, {\rm Tr}\, W(u) \Bigr).$$
Namely, we can compute
the superpotential by considering a deformation of the holomorphic 2-sphere
${\cal C_*}= \IP^1$ by giving $u$ a constant value on the $\IP^1$. This
deforms ${\cal C}_*$ to a nearby sphere ${\cal C}(u)$ which is not
holomorphic.
Then the superpotential is given by \refs{\wb,\AVS}
$$W(u) = \int_{B({\cal C}(u),{\cal C}_*)} \Omega.$$
We find\foot{As explained in more detail in \AV\ 
one can simplify the calculation by using independence
of the three-form periods on blowing up the geometry, which is a K\"ahler
deformation, and compute the integral in the singular geometry.
At fixed value of $u$, $\int dxdv/x$ integral computes
the holomorphic volume of the
special Lagrangian ${\bf S}^2$ in the two-fold fiber, and this is $u$.}
\eqn\suppb{W(u) = \frac {1}{2} u^2,}
as claimed above.

The state $|0_v\rangle$ can be found as follows.
In the context of a single D-brane, this is a simple $\delta$-function at 
$v=0$ since we have a non-interacting theory.
That is, we have
$|0_v\rangle=
\int {\cal D}v{\cal D}u \exp(\frac{1}{g_s}
\int_{\frac{1}{2}\IP^1} v^{(1)}{\bar \partial} u) $
which integrating over $v$ reduces to zero modes of $u$ and so
$$|0_v\rangle = \int du |u\rangle,$$
which is the same as in \dva.
More generally, for $N$ D-branes on the 
$\IP^1$ $u,v$ are promoted to matrices in the $U(N)$ Lie algebra, and
this will lead to non-trivial measure factors in the path integral written
in terms of eigenvalues.

Note that since $u,v$ are periodic in the geometry,
the natural measure for $N$ D-branes is not the 
Hermitian matrix measure as in \dva, 
but the unitary matrix measure, corresponding to a Hermitian
matrix with compact eigenvalues.
That is, for example in the B-model mirror to ${\bf S}^2\times {\bf S}^1$ 
we have 
\eqn\one{ \langle 0_v| 0_v\rangle = \frac{1}{{\rm vol}(U(N))}\int d_{H} u =  \frac{1}{|{\cal W}|} \int \prod_i d u_i
\;\Delta^2_{H}(u),}
where in the second equality we integrated over the angular variables of
matrix $u$ to get
$$\Delta_{H}(u) =\prod_{i<j} 2\sin(\frac{u_i-u_j}{2}).$$ 
This differs from the Hermitian matrix measure 
$\Delta(u) = \prod_{i<j} (u_i-u_j),$ and $\Delta_H (u)$ can be 
interpreted as a Hermitian measure in which 
we include the images of the D-brane \dvtwo, {\it i.e.} 
$$\Delta_H(u) \sim \prod_{n}  \prod_{i<j} (u_i-u_j+2 \pi n).$$ 
By taking the square root of \one, we find that 
$$|0_v\rangle = 
\frac{1}{|{\cal W}|^{1/2}} \int \prod_i d u_i
\Delta_H(u)|u\rangle,$$
where ${\bf u}_i |u\rangle = u_i |u\rangle$ is the eigenstate
of operator $u$. 

It is important to note that while $u$ and $v$ are periodic in $Y$,
the physics of the B-branes in these models generally does not 
have any periodicity, because the boundary conditions imposed
generally break this.
One can see this
already by considering a single D-brane in \mirrb. Taking $u$ to $u+2 \pi $
the ${\bf S}^2$ winds around the $v$ cylinder once: the south pole is
at $v=0$ and the north
pole at  $v=-u = -2\pi  $, and consequently the D-brane 
does not come back to itself. 
Alternatively, the superpotential is not periodic in $u$, and this 
corresponds to the fact that the tension of the D-brane increases in
going around. Consequently, the range of all integrations is non-compact.

The example of a B-brane on \mirrb\  
consequently gives a Hermitian matrix model, but with unitary measure
\eqn\paths{Z_{TST}=\langle 0_v | 0_{u+v} \rangle = 
\frac{1}{{\rm vol}(U(N))}\int d_H u 
\, \,e^{\frac{1}{2 g_s}{\rm Tr} u^2}.}
More general examples can be constructed along similar lines,
and we will see some of them in the following sections.

\ifig\circle{The B-brane projected to $v$ cylinder corresponds
to a path between $v=0$ and $v=-u$. Because the boundary
conditions on the two endpoints are different, going around $u \rightarrow
u=2\pi $, the B-brane does not come back to itself.}
{\epsfxsize4.0truein\epsfbox{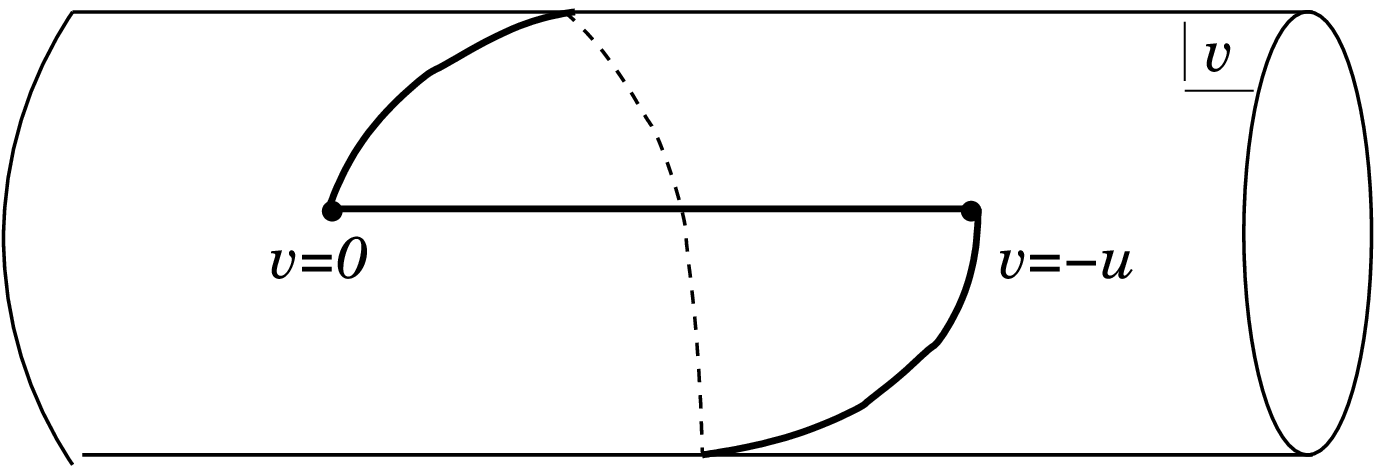}}

\newsec{Planar Limit}
In \gv\ it was shown that holes in the topological open string
amplitudes for $N$ D-branes on ${\bf S}^3$ in $X= T^*{\bf S}^3$
can be summed up, genus by genus. The resulting closed string amplitudes
coincided with that of closed topological A-model on 
$\hat X={\cal O}(-1)\oplus {\cal O}(-1)\rightarrow \IP^1.$
In the previous sections we showed that the Chern-Simons theory on 
${\bf S}^3$ 
can be rewritten as a matrix model 
\eqn\partb{Z = \frac{1}{{\rm vol}(U(N))}\int d_H u\  
\exp(\frac{1}{2 g_s} {\rm Tr} u^2)}
that naturally arises as the theory on the mirror B-model D-branes.
In this section we want to show that the matrix model is solvable in the
planar limit, and that the geometry which emerges is precisely that
of the mirror $\hat Y$ of ${\cal O}(-1)\oplus {\cal O}(-1) \rightarrow \IP^1$.

As discussed before, after integrating over angular variables, \partb\ 
can be written as
$$\int \prod_i du_i \Delta_H(u)^2 \exp(-{1\over 2g_s} \sum_i u^2_i).$$
Note that in writing the above integral we have made a choice of the
integration contour which amounts to $u_j \rightarrow i u_j$, and then  
$\Delta_H(u) = \prod_{i<j} 2 \sinh (\frac{u_i -u_j}{2})$. 
In the large $N$ limit, the integral is localized to
the saddle point,
\eqn\saddle{\frac{1}{g_s} u_i = \sum_{i\neq j} \coth(\frac{u_i - u_j}{2}),}
and we can replace the discrete set of eigenvalues $u_i$ by a 
continuous function $u(s)$. The sum in \saddle\ becomes an integral and 
we find
$$-\frac{1}{t} u(s_0) = {\rm P} \int_{0}^{1} ds \coth(u(s)-u(s_0)),$$
where $P$ denotes the principal value, and $t=N g_s$ is the 't Hooft parameter.
To solve the above equation we follow \bipz\ and 
introduce a density of eigenvalues $\rho(u)$. 
We now change variables from $u$ to $U = e^u$.
The density satisfies $\rho(U)dU/U = ds,$ and from $\int^{1}_{0}ds$
we have
\eqn\dens{ \int_{a}^{b} \rho(U) {dU \over U}= 1.}
In terms of $u$ the above equation \saddle\ is
\eqn\saddleb{ -\frac{1}{2t} \log(Ue^{-t}) = {\rm P} 
\int_{a}^{b}\frac{\rho(U')}{U'-U} dU',}
so solving \saddle\ is equivalent to solving
for the density of eigenvalues $u$ that satisfies \eqns{\dens,\saddleb}. 
The solution of \saddleb\ is now standard.
Namely, we can define a function $v(U)$ (usually called the resolvent) 
by
$$v(U) = t \int_{a}^{b} \frac{\rho(U')}{U'-U} dU',$$
and then the conditions on $u$ are equivalent to
asking that $(i)$ $v$ is analytic in the complex $U$ plane, 
cut along an interval $(a,b)$; $(ii)$ it decays at infinity as $1/U$;
$(iii)$ the period of $v$ around the cut is $2 \pi i t$; 
$(iv)$ as $U$ approaches the interval, $v(U\pm i \epsilon)= -\frac{1}{2} 
\log{(Ue^{-t})}\pm \pi i t \rho(U)$.

These conditions suffice to completely fix $v(U)$, to be
$$v = \log[\frac{1 + e^{-u} + \sqrt{(1+e^{-u})^2 - 4 e^{-u+t}}}{2}].$$
The zeros of the square root in the above expression correspond
to the endpoints of the cut. 
Alternatively, $v$ and $u$ are functions on the Riemann surface
\eqn\Riemann{(e^{v} - 1)(e^{v+u}-1)+e^{t} - 1 = 0,}
and moreover
there is a one-form $vdu$ whose periods on the Riemann surface
satisfy special geometry:
\eqn\pera{t = \frac{1}{2\pi i}\int_{A} v du,}
and
\eqn\perb{\partial_t F_0 = \frac{1}{2\pi i} \int_B v du,}
where A-cycle corresponds to integrating around the
cut, and the B-cycle corresponds to an integral from the endpoint of the
cut to some cut-off point at large $u$.

Note that on the one hand, the Riemann surface \Riemann\ is the non-trivial part of the geometry 
\eqn\mirrc{xz =(e^{v} - 1)(e^{v+u}-1)+e^{t} - 1,}
that arises by geometric transition that blows down the $\IP^1$ in \mirrb\ 
and deforms it by giving $t$ a non-zero value.
On the other hand, the equation \mirrc\ precisely describes the mirror of $ \hat{X} =O(-1)\oplus
O(-1) \rightarrow \IP^1$ \hv, where the size of the $\IP^1$ is mirror to 
$t$ in \mirrc.

In this sense, we have derived the mirror of $\hat{X}$ by showing the
equivalence of the open string A- and the B-model, and taking the
large $N$ limit of both.
It can be shown by explicit calculation that 
the function $F_0$ in \perb\ precisely agrees with
the genus zero partition function of the A-model on $\hat X$
and the sum over the planar diagrams in $U(N)$ Chern-Simons theory.

\newsec{Lens spaces}

In this section we consider a generalization of the above results 
where we replace
${\bf S}^3$ with the lens spaces $M_p = {\bf S}^3/\IZ_p$,
where $\IZ_p$ acts on ${\bf S}^3$ as
\eqn\lsp{|x|^2+|y|^2=1, \quad\;\;(x,y) \sim \exp(2i\pi/p) (x,y).}
We can think of this as obtained by gluing two solid 2-tori along their 
boundaries after performing the 
${\rm SL}(2,\IZ)$ transformation,
\eqn\lensmat{
U_p =\pmatrix{1 &  0 \cr
p & 1}.}
To see that, consider an ${\bf S}^3$ which, as explained above, is a $T^2$
fibration over an interval, where the cycles of the $T^2$
are generated by phases of $x,y$.
If the complex structure of the $T^2$ corresponding to ${\bf S}^3$
it is $\tau$, then an ${\rm SL}(2,{\bf Z})$ transformation that takes
this $T^2$ to a $T^2$ with $(1,0)$ and $(1,p)$ cycles vanishing
over the endpoints will take $\tau$ to $\tau'=\frac{\tau+1}{p}$.
But the $T^2$ with the new complex structure is precisely
a quotient of the original one by the ${\bf Z}_p$ action specified in
\lsp.

Wrapping $N$ D-branes on $M_p$ in $T^*M_p$, the topological A-model is
$U(N)$ Chern-Simons theory on $M_p$. 
The critical points of the CS action are flat connections, which are 
classified by embeddings of the first fundamental group in $U(N)$. 
Since $\IZ_p$ acts freely on ${\bf S}^3$, we have that $\pi_1(M_p) = \IZ_{p}$.
Therefore, on $M_p$ there are $\IZ_p$ discrete flat connections we can turn on.
A choice of a flat connection breaks the gauge group
$U(N)\rightarrow U(N_1)\times \ldots \times U(N_p)$, and leads to a choice
of vacuum of the theory. The full partition function of Chern-Simons theory 
on a compact manifold involves summing over all the flat connections, and 
in fact the nonperturbative answer that can be obtained from the relation
with WZW theory \jones\ gives such a sum. However, 
for our applications we are interested in Chern-Simons theory expanded
around a particular vacuum, so in evaluating Chern-Simons amplitudes
the prescription is not to sum over different flat connections.
Namely, although Chern-Simons theory lives in a compact space, in
our applications D-branes are wrapping not $M_p$ but $M_p\times \IR^4$,
corresponding to type IIA compactification on $T^*M_p \times \IR^4$. 

In this section we will first show how to generalize the B-matrix model to 
the case of lens space, and we will explicitly show that it agrees with the
direct computation using the standard techniques in CS. We will also 
discuss the large $N$ transition for CS on lens spaces, and we will
introduce a Hermitian multi-matrix model for CS 
that captures the contribution of a given vacuum.  

\subsec{B-model matrix model}

From the discussion in previous sections, the mirror
of $T^*M_p$ is given by blowing up
$$xy = (e^{v}-1)(e^{v+pu}-1),$$
corresponding to the fact that in the A-model, there are two lines
in the base $\IR^3$ over which the $(0,1)$ and $(p,1)$ cycles of the torus 
degenerate.
The resolved geometry,
$$xz = e^{v}-1,\quad u$$
in the $z-$patch and
$$yz' = e^{pu+v}-1, $$
in the $z'-$patch, $z=1/z'$.
There are $p$ holomorphic $\IP^1$'s at $v=0=pu,$
i.e. at
$$(u,v) = (2 \pi i k/p, 0), \quad k=0,\ldots, p-1.$$
Wrapping $N$ D-branes in this geometry, one has to decide how
to distribute the $N$ D-branes among the $p$ vacua. This we can see it 
at a quantitative level as well.
By a trivial generalization of \suppb, it is easy to see
that the theory on the wrapped D-branes
has a superpotential $W_p(u)$,
where
$$W_p(u) = p u^2/2,$$
and this has $p$ vacua as claimed.
The B-model path integral, as explained in section 3, is
\eqn\ls{Z = \langle 0_v|0_{v+pu}\rangle =\frac{1}{{\rm vol}(U(N))}\int d_H u \, {\rm e}^{-{1 \over g_s}
{\rm Tr} \, W_p(u)},}
since
$$\exp\Bigl({1 \over g_s}
{\rm Tr} \, W_p(u)\Bigr) : (u,v) \rightarrow (u, v+p u).$$
Distributing the $N$ branes among the $p$ different vacua corresponds, in 
the matrix model, to distributing the $N$ eigenvalues among the different 
critical points, and also to 
the choice of a flat connection 
in the Chern-Simons theory. 
 
Consider now the path integral
around the critical point
where
$N_{k}$ eigenvalues are at $u_j = 2\pi i (j-1)/p$, $\;j=1, \ldots p$,
and the gauge group is broken as $U(N) \rightarrow  U(N_1)\times 
\cdots \times U(N_{p})$. 
In the eigenvalue basis, the matrix model reads:
\eqn\lensmm{Z=
\int \prod_{j=1}^{p} {d^{N_j}u^{(j)}\over N_j!}  
\Delta_H(u^{(j)})^2
\prod_{j<k} \Delta_H(u^{(j)},u^{(k)})^2 \, \exp \Bigl\{ 
-\sum_j {\rm Tr}\,  p (u^{(j)})^2/2g_s \Bigr\} }
where we have denoted by $u^{(j)}$ the set of $N_i$ eigenvalues sitting at 
$2\pi i (j-1)/p$, and
$$
\eqalign{
\Delta_H(u^{(j)}) = &\prod_{m<n}2 \sinh\bigl({u^{(j)}_m - u^{(j)}_n \over 2}\bigr),\cr
\Delta_H(u^{(j)},u^{(k)}) =& \prod_{m,n} 2\sinh \bigl(
{u^{(j)}_m - u^{(k)}_n +d_{jk}\over 2}\bigr),\cr}$$
where $d_{jk}= 2 \pi i (j-k)/p$.
In other words, there is an effective interaction between D-branes
at different vacua. This can be thought of as coming  
from integrating out at one loop the massive string states stretched 
between the branes.

\subsec{Chern-Simons theory on ${\bf S}^3/\IZ_p$}

In this subsection we show that there is an exact agreement between
the topological B-model and the Chern-Simons answer, as expected.   
To do that, we 
will rewrite the matrix model \lensmm\ in the eigenvalue basis in a slightly
different way. Consider the integral
\eqn\integral{
\int\prod_{k=1}^N d u_k \, {\rm e}^{-\sum_j u^2_j/2
\hat g_s
-\hat k \sum_j n_j u_j} \prod_{j<k} \Bigl( 2 \sinh {u_j - u_k\over 2}
 \Bigr)^2,
}
where the effective coupling constant $\hat g_s$ is given by
\eqn\hatx{
\hat g_s = { 2\pi i \over p \hat{k}}.
}
In \integral, we have also introduced a vector $n$ of $N$ integer numbers
$0\le n_j \le p-1$ that label at which critical point is the
eigenvalue $u_j$. These integers label the choice of vacuum 
$U(N) \rightarrow U(N_1) \times \cdots \times U(N_p)$ as
follows: $N_k$ is
the number of $n_j$'s equal to $k-1$. Notice that there is not a one-to-one 
correspondence between the $n_j$'s and the different vacua, since any
Weyl permutation of the $n_j$ gives the same $N_k$'s. Therefore, there are in
total 
\eqn\deg{
{N! \over \prod_{k=1}^p N_k!}
} configurations of $n_j$'s that correspond to the same vacuum. 
Notice however that
\integral\ is manifestly invariant under
permutations of the $n_j$'s, so we can just pick any one of them. 
If we now change variables in \integral\ by $u_j \rightarrow 
u_j+\hat k \hat g_s n_j$, we reproduce \lensmm. 

According to our general results, the integral \integral\ must be the 
contribution of the flat connection labeled by $\{N_k\}_k$ to the partition
function of CS theory on $M_p$. This follows indeed from \mm, 
but in the case of lens spaces one can prove it in a very simple way. 
After using 
Weyl's denominator formula, the integral \integral\ becomes 
just a Gaussian, and it can be computed to give (up to overall constants)
\eqn\doublesum{
{1 \over | {\cal W}|} \sum_{w', w'' \in {\cal W}} \epsilon (w')\epsilon(w'') \exp
\Bigl\{ {i \pi \over \hat k  p} ( w'(\rho) - \hat k n - w''(\rho))^2 \Bigr\}.}
If we now sum \doublesum\ over all possible $n$, we obtain the following 
expression
\eqn\partfunc{
\sum_{n \in \IZ^N/p \IZ^N}
\sum_{w \in {\cal W}} \epsilon (w) \exp
\Bigl\{ {i \pi \over \hat k  p} ( \rho^2 - 2\rho\cdot(\hat k 
 n +w(\rho)) + (\hat k n +w(\rho))^2
)\Bigr\},}
To see this, notice that the lattice $\IZ^N/p \IZ^N$ in 
\partfunc\ is invariant under Weyl permutations, 
therefore we can sum over all possible permutations of $n$ and divide by the
order of the Weyl group $| {\cal W}|$. In this way we obtain \doublesum,
summed over all $n$. In this way, we have rederived the matrix element
\matsl\ when 
$U_p$ is 
the ${\rm SL}(2, \IZ)$ element \lensmat. Since this matrix element is the 
partition function of CS theory on the lens space $M_p$, we have shown that 
the integral \lensmm\ gives precisely 
the contribution of the flat connection labeled
by $\{N_k\}_k$  to the CS partition function. After including all the 
overall factors carefully, one finds that the precise expression of the
full partition function in the
canonical framing is
\eqn\intdef{
\sum_{n}   { {\rm e}^{- { \hat g_s \over 12}N(N^2-1)}\over N!}
\int\prod_{i=1}^N {d u_i \over 2\pi} \, {\rm e}^{-\sum_i u^2_i/2
\hat g_s
-\hat k \sum_i n_i u_i} \prod_{i<j} \Bigl( 2 \sinh {u_i - u_j\over 2}
 \Bigr)^2.
}

\subsec{Large $N$ duality for lens spaces}
In \gv, the large $N$ limit of topological open strings on $T^* {\bf S}^3$ 
was shown to be given by closed topological strings on the resolved
conifold ${\cal O}(-1) \oplus {\cal O}(-1) \rightarrow \IP^1$.  
There is a natural question
of what is the large $N$ limit when we replace ${\bf S}^3$ with 
${\bf S}^3/\IZ_p$. The answer for this, generalizing \gv, is as follows.
For definiteness, consider first $p=2$. As is familiar,
$X= T^*({\bf S}^3/\IZ_2)$ has a geometric transition where
${\bf S}^3/\IZ_2$ is replaced by $F_0= \IP^1\times \IP^1$.
The total geometry is a cone over this, more precisely
it is $\hat X = O(-K) \rightarrow F_0$.
\ifig\lens{The figure depicts a geometric transition between
$T^*({\bf S}^3/\IZ_2)$
and $O(-K)\rightarrow \IP^1 \times \IP^1$.
With $N$ D-branes on ${\bf S}^{3}/\IZ_2$ and gauge group broken
to $U(N_1)\times U(N_2)$ with $N=N_1+N_2$, the geometric transition is a
large $N$ duality and the BPS sizes $S_{1,2}$ of two $\IP^1$'s are
identified with the t'Hooft parameters $S_i = N_i g_s$.}
{\epsfxsize5.5truein\epsfbox{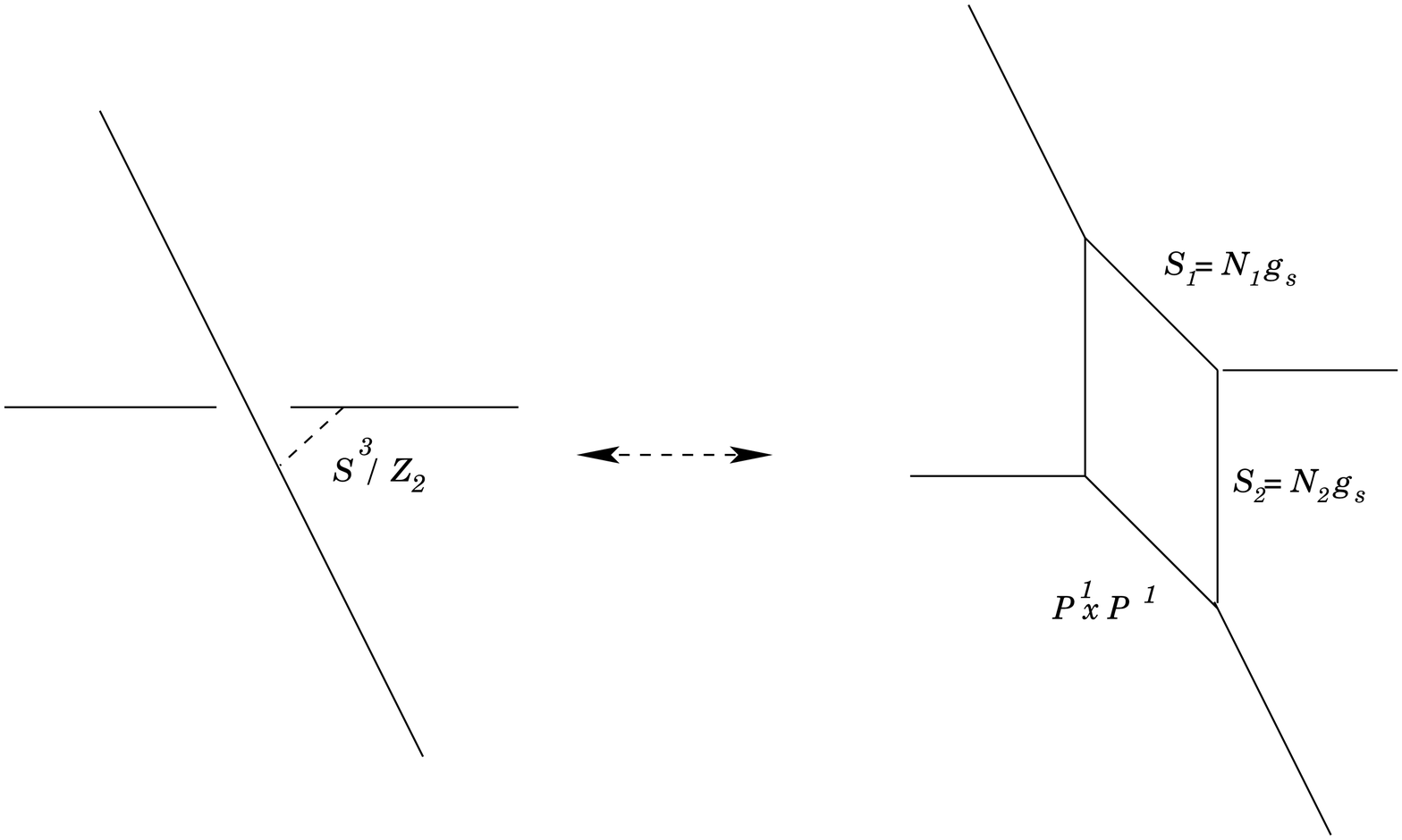}}
That such transition is allowed is easy to see
in the language of $(p,q)$ five-branes, see \lens.
One may then expect that the large $N$ limit of $N$
D-branes on ${\bf S}^3/\IZ_2$ is a closed string theory on $\hat X$. For 
general $p$, the dual geometry is an $A_{p-1}$ fibration over $\IP^1$, 
with $p$ complexified K\"ahler classes corresponding to the sizes of the 
$p$ different $\IP^1$'s.

In order to make precise the implications of this large $N$
transition, we need an identification of the parameters
between the two theories. On the open string side we have
a choice of the numbers $N_k$ of D-branes to place in
the $p$ different vacua, and we would expect that these choices correspond to 
changing the sizes of the $p$ $\IP^1$'s. The natural identification is 
as follows.

Recall that in the open string theory 
the large $N$ expansion is a weak coupling expansion in
$g_s$. The open string free-energy is of the form
$$F =F^{\rm nonpert} + F^{\rm pert},$$
where
$$
F^{\rm pert}=\sum_{g=0}^{\infty}F^{\rm pert}_{g,h}(N_k) g_s^{2g-2+ h}.$$
The $N_k$ dependence in $F^{\rm pert}_g$ comes from tracing over
the Chan-Paton indices of Riemann surfaces with holes. This expansion is
nothing but the Feynman-diagram expansion of the CS path integral in the
background of a flat connection given by the $N_k$'s. Notice that $F^{\rm
pert}_{g,h}(N_k)$ has in fact the structure
\eqn\pertstr{
F^{\rm pert}_{g,h}(N_k)=\sum_{h_1+ \cdots +h_p=h}F^{\rm pert}_{g, h_1, \cdots, h_p} 
N_1^{h_1} \cdots N_p^{h_p},}
{\it i.e.} it is a homogeneous polynomial in $N_k$ of degree $h$. 
The non-perturbative piece, in contrast to the model 
dependence of $F^{\rm pert}$, has a 
universal behavior. From the open string/CS perspective, this
comes from the measure of the path integral -- basically the volume of 
the unbroken gauge group $G$ \ov. 
$$F^{\rm nonpert} \sim - \log ({\rm vol}(G)).$$
In our case $G= 
U(N_1) \times \cdots \times U(N_p)$, and the explicit expression of $F^{\rm
nonpert}$ can be easily obtained from the asymptotic expansion 
$$
\eqalign{
\log ({\rm vol}(U(N)))=&-{N^2 \over 2} \Bigl( \log (N) -{3 \over 2} \Bigr) 
+{1 \over 12}\log N + {1 \over 2}N^2 \log 2 \pi \cr
& - \sum_{g=2}^{\infty} {B_{2g} \over
2g (2g-2)} N^{2-2g}.\cr}
$$
In order to identify the parameters in the open and the closed
string side, consider the genus zero piece of the nonperturbative part of the 
free energy:
\eqn\leadingf{F^{\rm nonpert}_{g=0}  = \frac{1}{2} \sum_{i=1}^{p} (g_s N_i)^2 
\log(g_s N_i) ,}
where $g_s N_i$ is the 't Hooft coupling. 
This universal behavior strongly suggests the following.
The genus zero topological closed string amplitudes with  
the above form are well known to arise by integrating
out nearly massless charged particles of mass $g_s N_i$, since 
\leadingf\ is basically a contribution of
BPS D-branes at one loop to the ${\cal N}= 2$ prepotential. Therefore, one may 
naively identify the 't Hooft parameters $g_s N_i$ with the flat coordinates
$S_i$ measuring the BPS sizes of the $p$ $\IP^1$'s: 
\eqn\matchvar{
S_i =g_s N _i, \,\,\,\,\,\,\,\, i=1, \cdots, p.} 
Notice that in this picture the perturbative piece of the 
open string free energy
$$F^{\rm pert}_g (S_i)=\sum_{g, h_1, \cdots, h_p} F^{\rm pert}_{g,h_1,
\cdots, h_p} (g_sN_1)^{h_1} \cdots (g_s N_p)^{h_p}$$
which can be computed in ordinary Chern-Simons perturbation theory, 
is the regular part 
of the $F^{\rm closed}_g$ coupling for the closed string 
dual geometry, expanded in
terms of flat coordinates $S_i$ around the point in moduli space where the
$\IP^1$'s have vanishing size. We will refer to this point as the orbifold
point, although in contrast to orbifold points in other
geometries, like local $\IP^2$, we have a singular behavior of the
prepotential captured by $F^{\rm nonpert}$. 
We will show below that the naive
expectation \matchvar\ is correct, 
by comparing the perturbative expansion in the open
side with the expansion of $F^{\rm closed}_g$ computed in the B model
around the orbifold point. 
\subsec{(Hermitian) Matrix model for Chern-Simons on lens spaces}

In order to test the large $N$ duality for lens spaces in the way that we
just suggested, we have to compute $F_{g,h}$ in the open string/CS side. 
To do this the equivalence between CS theory 
and matrix models turns out to be very useful. As pointed out in \mm, 
one can regard the CS matrix model as a 
``deformation" of the usual Hermitian Gaussian model, where the deformation
is due to the appearance of $\prod_{i<j} (2 \sinh((u_i -u_j)/2))^2$ 
instead of the usual Vandermonde determinant $\prod_{i<j} (u_i
-u_j)^2$, therefore one can systematically compute the
perturbative expansion of the CS theory in terms of perturbation theory of
the gauged matrix model around the Gaussian point. 

We will in fact 
write a {\it Hermitian} matrix model underlying \intdef.
Let us first consider the contribution due to the trivial connection, {\it i.e.}
let us consider the integral in \intdef\ with $n=0$. 
We now do the following trick. As in \mm, we write
\eqn\decomp{
\prod_{i<j} \Bigl( 2 \sinh {u_i - u_j\over 2}
 \Bigr)^2 = {\Delta}^2 (u) f(u).
}
In this equation ${\Delta} (u)=\prod_{i<j} (u_i - u_j)$
is the usual Vandermonde determinant, and the function $f(u)$ is given by
\eqn\fb{
f(u)= \exp\biggl( \sum_{k=1}^\infty a_k \sigma_k (u)\biggr),}
where
\eqn\sig{
\eqalign{
\sigma_k(u)=&\sum_{i<j} (u_i -u_j)^{2k},\cr
a_k=&{ B_{2k} \over k (2k)!}\cr}
}
and $B_{2k}$ are the Bernoulli numbers. $a_k$ are simply the
coefficients in the expansion of $\log (2 \sinh(x/2)/x)$. The
$\sigma_k(u)$ are symmetric polynomials in the $u_i$'s, therefore
can be written in terms of Newton polynomials
\eqn\new{
P_j(u) =\sum_{i=1}^N u_i^j,}
as follows:
\eqn\signew{
\sigma_n (u)= NP_{2n}(u) + {1 \over 2} \sum_{s=1}^{2n-1}
(-1)^s {2n \choose
s} P_s(u) P_{2n-s}(u).}
We then write the integral as:
\eqn\mmher{
{ {\rm e}^{- { \hat g_s \over 12}N(N^2-1)}\over N!}
\int\prod_{i=1}^N {d u_i \over 2\pi}
{ \Delta}^2(u)\, \exp \biggl( -\sum_i u^2_i/2
\hat g_s + \sum_{k=1}^\infty a_k \sigma_k (u)\biggr).}
Now we notice that the
Newton polynomials $P_j(u)$ are just ${\rm Tr}M^j$, where $M$ is a
Hermitian matrix which has been gauge-fixed to the diagonal form ${\rm
diag}(u_1, \cdots, u_N)$. Therefore the above
integral is (up to the prefactor ${\rm e}^{- { \hat g_s \over 12}N(N^2-1)}$)
the gauge-fixed version of the Hermitian matrix model
\eqn\herm{
{1 \over {\rm vol}(U(N))}\int dM \exp \biggl\{-{1 \over 2 \hat g_s} {\rm
Tr}M^2 + V(M) \biggr\},}
where
\eqn\espm{
V(M)=
{1 \over 2}\sum_{k=1}^{\infty} a_k \sum_{s=0}^{2k}
(-1)^s {2k \choose
s} {\rm Tr}M^s {\rm Tr}M^{2k-s}.}
Here we used the expression for the Hermitian measure
(see for example the second appendix in \biz)
\eqn\meas{
{1 \over {\rm vol}(U(N))} dM = {1 \over N!} {\Delta}^2(u)
\prod_{i=1}^N du_i,}
up to factors of $2$ and $\pi$.
Therefore in \espm\
we have represented the eigenvalue interaction of \integral\ in
terms of an infinite number of vertices. Notice however that,
at every order in $\hat g_s$, only a finite number of vertices
contribute, so the perturbation expansion $\sum_{g,h} F_{g,h} \hat g_s^{2g-2+
h} N^h$ of the partition function can be computed from the Hermitian matrix
model \herm\ with action \espm. In order to obtain the
perturbation expansion of \herm,
we just bring down the powers of ${\rm Tr}M^j {\rm Tr}M^k$ from the
exponent and we evaluate the vevs with the Gaussian weight $\exp(
 -{1 \over 2 \hat g_s} {\rm Tr}M^2)$. The Gaussian averages can be computed in
many ways, and we review some of these techniques in Appendix A.

Let us now consider the expansion around
a nontrivial flat connection, focusing on $p=2$ (the
general case is similar). The resulting integral is given 
by \lensmm\ with $p=2$. Equivalently, we can obtain it 
by expanding around the critical point $u^*=-i\pi  n$ of
the exponent in \integral. We will take
the representative of $n$ in the Weyl orbit given by
\eqn\trep{
n=(0, \cdots, 0, 1, \cdots, 1)
}
where there are $N_1$ $0$'s and $N_2$ 1's. There are two groups of 
integration variables, as in \lensmm, that we will denote by 
$\{ \lambda_i\}_{i=1, \cdots, N_1}$, and
$\{ \mu_i\}_{i=1, \cdots, N_2}$. The measure factor in \lensmm\ reads now:
\eqn\calmeas{
(-1)^{N_1 N_2} \prod_{1\le i<j\le N_1}\Bigl( 2 \sinh {\lambda_{i} -\lambda_{j} \over 2}
 \Bigr)^2 \prod_{1\le i<j\le N_2}
\Bigl( 2 \sinh {\mu_{i} -\mu_{j} \over 2}
 \Bigr)^2 \prod_{i,j}\Bigl( 2 \cosh {\lambda_i -\mu_j\over 2}
 \Bigr)^2.}
The model is then equivalent to a two-matrix model with an $N_1 \times N_1$
Hermitian matrix $M_1$ and an $N_2 \times N_2$ Hermitian 
matrix $M_2$. The two matrices
interact through the last factor in \calmeas, that can be written as:
\eqn\interac{
\exp \biggl\{ 2 \sum_{i,j} \log \Bigl( 2 \cosh {\lambda_i -\mu_j\over 2}
\Bigr) \biggr\}.}
In terms of $M_1$ and $M_2$, this is
\eqn\hermint{
W(M_1, M_2) = \sum_{k=1}^{\infty} b_k \sum_{s=0}^{2k}
(-1)^s {2k \choose
s} {\rm Tr}M_1^s {\rm Tr}M_2^{2k-s},}
where
\eqn\bk{
b_k= {2^{2k}-1 \over k (2k)!}B_{2k}.}
On the other hand, $M_1$ and $M_2$ interact with themselves through the
potentials $V(M_1)$, $V(M_2)$, given in \espm. Making use of \meas\ we
finally obtain an ``effective'' two-matrix model given by:
\eqn\efftwo{
\eqalign{
& {1\over {\rm vol}(U(N_1))
\times {\rm vol}(U(N_2))}  \cr
& \times \int dM_1 dM_2 \exp \biggl\{ -{1 \over 2 \hat g_s} {\rm
Tr}M_1^2 -{1 \over 2 \hat g_s} {\rm
Tr}M_2^2+ V(M_1)+ V(M_2) + W(M_1, M_2)\biggr\}.\cr}
}
Similar ideas and techniques to analyze matrix models expanded around
nontrivial vacua have been presented in \dgkv\ (see also \bd).
 
In \efftwo\ we have omitted an overall factor:
\eqn\class{
(-4)^{N_1 N_2} {\rm e}^{- { \hat g_s \over 12}N(N^2-1)}
{\rm e}^{\hat k  N_2 \pi i  /2}
,}
where the last factor equals
$\exp \Bigl\{ {1 \over 2 \hat g_s} (u^*)^2 \Bigr\}$,
which is the value of the classical CS action on the flat connection
associated to \trep. Notice that the overall factor $
{ {\rm e}^{\hat k  N_2 \pi i  /2 }\over {\rm vol}(U(N_1)) \times {\rm
vol}(U(N_2))}$ is in agreement with the prediction for the structure of the
semiclassical expansion of CS \roz.

Using \efftwo, the perturbative expansion around the nontrivial
flat connection is just a matter of computing averages in the Gaussian
ensemble. We have computed the perturbative 
free energy 
$F^{\rm pert}=\sum_g F^{\rm pert}_{g,h}(N_1,
N_2) 
\hat g_s ^{2g-2 + h}$ up to order $4$ in the 
effective coupling constant. These 
quantities are homogeneous, symmetric polynomials of degree $h$ in
$N_1$, $N_2$. For genus $0$ one has:
\eqn\res{
\eqalign{
F^{\rm pert}_{0,4}=&{1 \over 288} \Bigl\{ N_1 ^4 + 6 N_1^3 N_2 + 18 N_1^2 N_2^2
+ 6 N_1 N_2^3 + N_2^4 \Bigr\},\cr
F^{\rm pert}_{0,6}=&-{1 \over 345600} \Bigl\{ 4 N_1^6 + 45 N_1^5 N_2 + 225 N_1^4 N_2
^2
+ 1500 N_1^3 N_2^3 \cr
& \,\,\,\,\,\,\,\,\,\,\,\,\,\,\,\, + 225 N_1^2 N_2^4 + 45 N_1 N_2^5 + 4 N_2^6\Bigr\}.\cr}}
For genus $1$, one finds:
\eqn\resone{
\eqalign{
F^{\rm pert}_{1,2}=&-{1 \over 288} \Bigl\{ N_1 ^2 - 6 N_1 N_2 +  N_2^2 \Bigr\},\cr
F^{\rm pert}_{1,4}=&{ 1\over 69120} \Bigl\{ 2 N_1^4 + 105 N_1^3 N_2
-90 N_1^2 N_2^2
+ 105 N_1 N_2^3 + 2 N_2^4 \Bigr\}.\cr}}
Finally, for genus $2$ one finds:
\eqn\restwo{
F^{\rm pert}_{2,2}=-{1 \over 57600} \Bigl\{ N_1^2 + 60 N_1 N_2 + N_2^2 \Bigr\}.
}
As a partial check of these expressions, notice that, if $N_1=N$ and 
$N_2=0$ ({\it i.e.} when we specialize to the trivial connection) the 
partition function of $M_p$ is identical to the partition function 
on ${\bf S}^3$, up to a rescaling of the coupling constant, and  
the coefficients $F^{\rm pert}_{g,h}(N)$ can be obtained from the results of 
\gvmone\gv. Their explicit expression is
\eqn\fghsimple{
\eqalign{
F^{\rm pert}_{0,h}=& {B_{h-2} \over (h-2) h!} \cr 
F^{\rm pert}_{1,h}=&-{1 \over 12}{ B_h \over h\, h!}   \cr 
F^{\rm pert}_{g,h}=&-{1 \over h!} {B_{2g-2 + h} \over 
2g-2 + h}{B_{2g} \over 2g (2g-2)},\,\,\,\, g \ge 2,\cr}}
in agreement with the above results for $N_2=0$.
In the next sections we will see that the above expansions 
exactly agree with the expansion of the closed string amplitudes on 
local $\IP^1 \times \IP^1$ near the orbifold point.

For ${\bf S}^3/\IZ_p$ with general $p$, the result for an arbitrary
flat connection can be written as a $p$-matrix model
\eqn\generalp{
\eqalign{
& {1\over \prod_{i=1}^p {\rm vol}(U(N_i))}  \cr
& \times \int \prod_{i=1}^p dM_i \exp \biggl\{ -{1 \over 2 \hat g_s}
\sum_{i=1}^p {\rm
Tr}M_i^2 + \sum_{i=1}^p V(M_i) + \sum_{1\le i<j \le p} W(M_i, M_j)\biggr\},\cr}
}
where $V(M)$ is still given by \espm, and $W(M_i, M_j)$ is
given by
\eqn\wgen{
W(M_i, M_j)=\sum_{k=1}^{\infty} 2^{-k+1}a_k^{(ij)}\sum_{s=0}^{k}
(-1)^s {k \choose
s} {\rm Tr}M_1^s {\rm Tr}M_2^{k-s},}
and $a_k^{(ij)}$ are
the coefficients in the Taylor series expansion of
\eqn\serlog{
\log \sinh \Bigl( (j-i){ \pi i \over p} + x \Bigr)
.}

\newsec{Closed topological strings on ${\cal O}(-K)\rightarrow \IP^1 \times \IP^1$.}

In this section we will calculate the topological 
string amplitudes for the non-compact Calabi-Yau geometry
which is the large $N$ dual of 
$T^{*} {\bf S}^3/\IZ_2$, by using
mirror symmetry and the B-model technique. 
The geometry is the canonical line 
bundle over $F_0=\IP^1\times \IP^1$. 
The B-model mirror
description of that geometry is encoded in a Riemann surface with 
a meromorphic differential. 
Many of the techniques developed here extend to more general
non-compact Calabi-Yau geometries.

\subsec{Moduli space of ${\cal O}(-K)\rightarrow F_0$}

Let us first describe the complexified  
K\"ahler moduli space. 
\ifig\unresolved{Schematic view of the unresolved moduli space of
${\cal O}(-K) \rightarrow \IP^1\times \IP^1$.}
{\epsfxsize 3.8truein\epsfbox{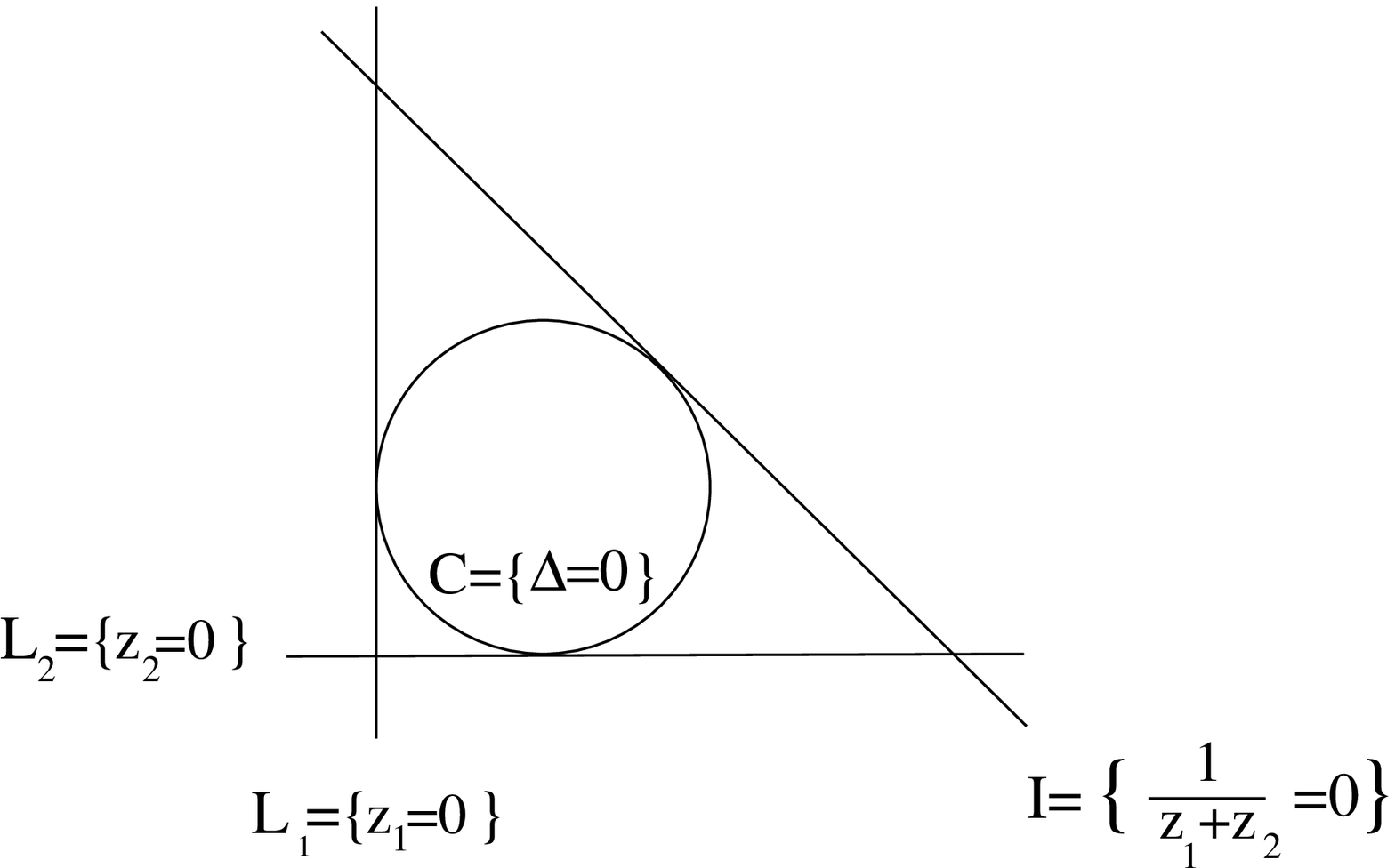}}
The method  to analyze the complexified K\"ahler moduli space 
is to study the complex structure deformations of the mirror
as encoded in the period integrals. Up to finite choice of 
integration constants these periods are captured by the linear differential
operators of order two \kkvI \foot{The Picard-Fuchs equations as 
starting point 
of the further considerations can be easily obtained for all toric non-compact Calabi-Yau
spaces \ckyz. Using the mirror geometry given most explicitly in 
\hv, it would be also possible to work
directly with period integrals.}
\eqn\pfpIpI{\eqalign{
{\cal L}_1 &= z_2 (1-4 z_2) \xi^2_2 - 4 z_1^2 \xi_1^2 -
8 z_1 z_2 \xi_1 \xi_2 - (6 z_1+ 6 z_2)\xi_1 + \xi_2, \cr
{\cal L}_2 &= z_1 (1-4 z_1) \xi^2_1 - 4 z_2^2 \xi_2^2 -
8 z_1 z_2 \xi_1 \xi_2 - (6 z_1+ 6 z_2)\xi_2 + \xi_1, }}
where the $\xi_i={\partial\over \partial z_i}$. Differential
systems governing the periods can have only regular singular
points \arnoldbookII , i.e. the periods will in ``suitable''
coordinates have at worst  (in this case double) logarithmic
singularities. One can obtain the corresponding
singular locus by calculating the resultant of
the leading (order two) pieces of ${\cal L}_i=0$
with $\xi_i$ viewed as algebraic variables. This yields
$$z_1 z_2 [1-8(z_1+z_2)+16 (z_1-z_2)^2]=:z_1 z_2\Delta=0\ . $$
We need to compactify the $z_1,z_2$ space and chose $\IP^2$ as
first approximation to do that, i.e. we  consider in addition
the patches $(a_1=1/z_2,a_2=z_1/z_2)$ and $(b_1=1/z_1,b_2=z_2/z_1)$.
Transforming \pfpIpI\ and repeating the analysis in these coordinates we get the
following schematic picture of the degeneration locus in \unresolved. 
We see that the $C$ touches $L_1$ at $z_2={1\over 4}$, $L_2$ at
$z_1={1\over 4}$ and $I$ at $u={z_1\over z_1 +z_2}={1\over 2}$. All intersections are with contact 
order two. For example identifying\foot{Similarly at $C\cap L_2$ we set
$a=(1-4 z_1)$ and $b=z_2$ and at $C\cap L_1$,  $a=(1-4 z_2)$ and $b=z_1$.}
at $C\cap I$ $a=4 (1 - 2 u)$ and $b={8\over {z_1+z_2}} $ the local equations 
at the intersection $C\cap I$ 
are
\eqn\localsing{C=\{a^2-b=0\}\ \ \ {\rm and} \ \ I=\{b=0\}\ . }
As a consequence the differential equations are not solvable in the 
local variables $(a,b)$. Physically speaking we have to consider a 
multi scaling limit in approaching the intersection point in order 
to be able to define the $F^{(g)}$. 
\ifig\resolved{Schematic view of the resolved moduli space of
${\cal O}(-K) \rightarrow \IP^1\times \IP^1$.}
{\epsfxsize 3.5truein\epsfbox{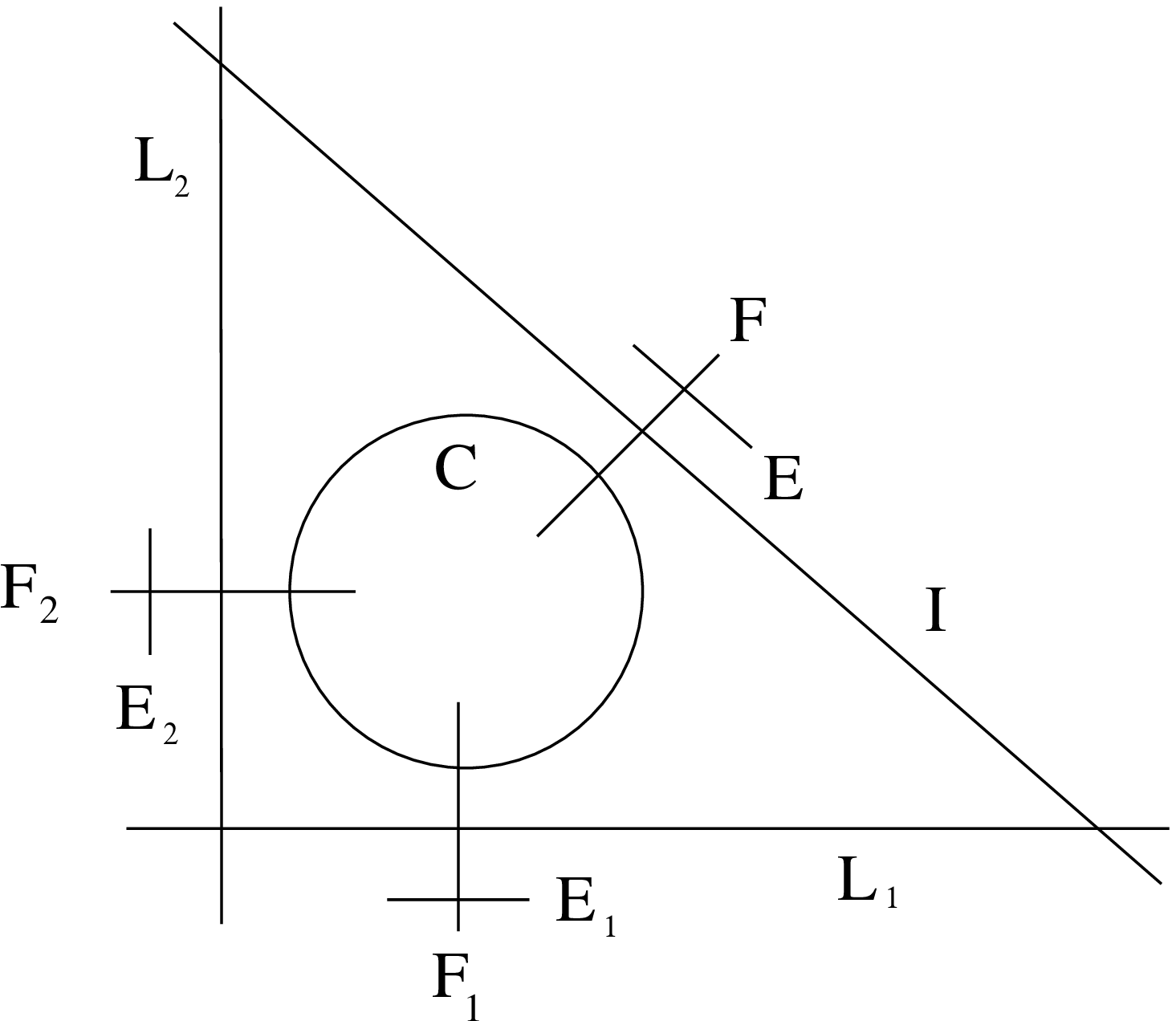}}

In algebraic geometry this corresponds to the 
well-known fact that one can resolve the moduli space of Kuranishi family 
in a way that all boundary divisors, i.e. the discriminant components, 
have normal crossings. The vanishing coordinates at those divisors are the 
``suitable'' coordinates for the statement about the regular singular 
behavior of the periods above. The resolution process of \localsing\ 
is standard and was used in similar context in \CandelasDM . To 
resolve points of contact order $k$ one introduces
$k$ times $(a_i:b_i)$  homogeneous $P^1$ variables and $k$ relations. 
In our case the process produces normal crossing after 
introducing  $a_1 a = b_1 b$  and  $a_2 a=b_2 a_1$. In the 
$(a,b|a_1:b_1|a_2:b_2)$ variables the coordinates along the divisors
are $C$: $(\sqrt{b},b|a:1,|1:1)$, $I$: $(a,0|0:1|0:1)$, $E$:
$(0,0|a_1:1|1:0)$ and finally $F$: $(0,0|0:1|a_2:b_2)$. One sees
from that that $F$ intersects $I$, $E$ and $C$ in the way depicted 
in \resolved. The blow ups of $L_i\cap C$ are completely analogous 
resolutions of the local equation \localsing.

For us the most relevant points are
$$\eqalign{
I\cap F: 
\quad \IZ_2 {\rm  \, orbifold\ pt.}: \quad {\rm Matrix\ model\ expansion,}
\cr
L_1\cap L_2:
\quad {\rm large\ complex\ st. \ pt.}: \quad  {\rm Topological\ A-model\ 
expansion. }}$$

Also interesting are the two copies of Seiberg-Witten field theory embedding
$$\eqalign{
L_i\cap F_i: 
\quad {\rm  SW\ weak\ coupl.\ pt.}: \quad {\rm Space\ time\ instanton\ expansion, }
\cr
L_i\cap C:
\quad {\rm  SW\ strong\ coupl.\ pt.}: \quad  {\rm SW\ strong \ coupling\  expansion\ scheme.}}$$

\subsec{Choosing local complex structure coordinates}

Choosing local complex structure coordinates is
merely a technical issue needed to evaluate the periods at
all points, two of which will really become the good physical
B-model variables. The transversal directions to the divisors are the
good complex coordinates coordinates. At $I\cap F$, $b_2=1$
and $a_2$ moves transversally to $I$ along $F$ so $a_2=a_1/a={b\over a^2}$
and $a$ moves transversally to $F$ along $I$, i.e. $(b/a^2,a)$ are
good coordinates. At $F\cap E$ $b_2={a^2\over b}$ is
transverse to $E$ and $a_1={b\over a}$ transverse to $F$, good
coordinates are $({a^2\over b},{b\over a})$ and finally at $C\cap F$.
And at $C\cap I$: $(1-{b\over a^2},a)$ are good coordinates. This
clarifies the choice of the complex structure variables at all blown loci. At
$L_1\cap L_2$ good local variables are $(z_1,z_2)$ and at $L_i\cap I$
$\left({z_i\over z_1+z_2},{1\over z_1+z_2}\right)$. 
Clearly the right choice of these variables is a local issue,
e.g. we could also have chosen  $\left({z_1\over z_2},
{1\over z_2}\right)$ at  $L_1\cap I$ which differs only
away from $L_1$ from the previous ones.

A global issue in the choice of complex parameters is the fact that  
$(z_1,z_2)$ are actually $\IZ_2\times \IZ_2$ multi covering variables. 
The branching loci of which give rise to  the $E$-type divisors. 
Choosing single cover variables $x_1^2=z_1$ and $x_2^2=z_2$ the conifold locus $\Delta=0$ 
reduces into four components and the embedding of the Seiberg-Witten $u$-planes ($F_1,F_2$
in \resolved.) become more familiar since there is now a $(1,-1)$ dyon 
component and a $(0,-1)$ monopole component crossing the four $u$-planes in the 
single cover variables.

\subsec{Solving the Picard-Fuchs equation near the orbifold point}

In the usual application of mirror symmetry the periods are evaluated near 
the large complex structure point $L_1\cap L_2$. Two of the periods, usually called 
$t_1=\log(z_1)+{\cal O}(z)$ and $t_2=\log(z_2)+{\cal O}(z)$, approximate at 
this point the classical large K\"ahler volumes of the two $\IP^1$.  

Here we need to expand the solutions to the Picard-Fuchs equations near 
the orbifold point. It is convenient to use the variables
\eqn\cov{\left(x_1=1-{z_1\over z_2}, x_2={1\over \sqrt{z_2}\left(1-{z_1\over z_2}\right)}\right)\  . }
The choice of $x_1$ and (B.5) ensures that $q_1=q_2$ or $t_1=t_2$ near the
expansion point, while the vanishing of $x_2$ ensures that $\sqrt{z_2}$ goes faster to infinity
then $x_1$ goes to zero. 

The periods in this variables have the following structure 
$$\eqalign{
\omega_0&=1,\cr
s_1&=-\log(1-x_1)=\sum_{m} c_{m,0} x_1^m=t_1-t_2,\cr
s_2&=\sum_{m,n} c_{m,n} x_1^m x_2^m,\cr
F^{(0)}_{s_2}&=s_2 \log(x_1)+\sum_{m,n} d_{m,n} x_1^m x_2^n \ ,}$$
where the $c_{m,n}$ and $d_{m,n}$ are determined by the following recursions relations
$$\eqalign{
c_{m,n}=&c_{m-1,n}{(n+2-2m)^2\over 4 (m-n) (m-1)},\cr
c_{m,n}=&{1\over n(n-1)} (c_{m,n-2} (n-m-1)(n-m-2)- c_{m-1,n-2} (n-m-1)^2),}
$$
$$\eqalign{
d_{m,n}=&{d_{m-1,n}(n+2-2m)^2 + 4 (n+1-2m) c_{m,n}+ 4 (2m-n-2) c_{m-1,n}\over 4 (m-n) (m-1)},\cr
d_{m,n}=&{1\over n(n-1)} (d_{m,n-2} (n-m-1)(n-m-2)- d_{m-1,n-2} (n-m-1)^2 \cr &+(2n-2-2m)c_{m-1,n-2}+(2m+3-2n)c_{m,n-2}).}
$$

Up to linear transformations we expect the $s_1$ and $s_2$ periods to be the good coordinates 
in which we will express the B-model correlators, which are giving in  $(x_1,x_2)$ 
coordinates using (B.3) and \cov . We therefore need the inverse function $x(s)$.
To invert the second and third period we define ${\tilde s}_1=s_1=x_1+{\cal O}(x^2)$ and 
$\tilde s_2={s_1\over s_2}=x_2+{\cal O}(x^2)$ 
\eqn\inversion{
\eqalign{x_1(s_1)&=1-e^{-\tilde s_1},\cr 
x_2({\tilde s}_1,{\tilde s}_2)&={\tilde s}_2+{1\over 4} {\tilde s}_1 {\tilde s}_2 + {1\over 192}  {\tilde s}^2_1 {\tilde s}_2  - 
    {1\over 256}   {\tilde s}^3_1 {\tilde s}_2 - {49\over 737280} {\tilde s}^4_1 {\tilde s}_2 - 
          {1\over 192} {\tilde s}^2_1 {\tilde s}^3_2+{\cal O} ({\tilde s}^6)\ . }} 
This yields the mirror map at the orbifold point.

\subsec{The genus zero partition function at the orbifold point}

The genus zero partition function can now be obtained by transforming (B.3) using \cov\ to
the $(x_1,x_2)$ and by \inversion\ to the $(s_1,s_2)$ coordinates. These $s$ variables
are flat coordinates, which have natural ${\rm GL}(2,\IC)$ structure.
It follows that we can integrate the $c_{ijk}(s)=\partial_{s_i}\partial_{s_j} \partial_{s_k} F^{(0)}$
to obtain the prepotential  $F^{(0)}$ up to a 
quadratic polynomial in $s$. The appropriate variables
$S_1,S_2$ that match the 't Hooft parameters in the CS/matrix model side 
are given by
\eqn\matrixmodelperiods{S_1={1\over 4}(s_1+s_2),\quad \quad  S_2={1\over 4}(s_1-s_2)\ . }
In view of these identifications, the fact that $s_1=t_1-t_2$ and the symmetry of $S_1,S_2$
in the partition functions below we conclude that $s_2=t_1+t_2$, hence $S_i={1\over 2} t_i$.
This can be shown also by analytic continuation.

An alternative way to get $F^{(0)}$ is to integrate $F^{(0)}_{s_2}$ with respect to the flat
coordinate $s_2$. This way one misses terms, which depend only on $s_1$, but those can be reinstalled
by requiring symmetry between $S_1$ and $S_2$ in the final expression. So one can get
$F^{(0)}$ up to a constant. By comparing the all genus partition function
$F=\sum_{i=0}^\infty g_s^{2g-2} F^{(g)}$ with the matrix model one 
also has to make
a choice of the string coupling $g_s$ namely $g_s^{\rm top}= 
2 i\hat g_s$.
This way the terms in front of $\hat g_s^{-2}$ are
\eqn\genuszero{F^{(0)}= {1\over 2} ( S_1^2\log(S_1)+S_2^2 \log(S_2) ) + \sum_{m,n} c^{(0)} S_1^m S_2^n + p_2(S)\ . }
The $c^{(0)}_{m,n}$ are only non-zero for $n+m\in 2 \ZZ$ 
and symmetric in $m,n$. The first few degrees
have been checked against the matrix model calculation in \res:
$$
\eqalign{
d=4: &\qquad {1\over 288}(S_1^4 + 6 S_1^3 S_2 + 18 S_1^2 S_2^2 + 6 S_1 S_2^3 + S_2^4)\cr
d=6: &\qquad -{1\over 345600 }(4 S_1^6 +45 S_1^5 S_2 + 225 S_1^4 S_2^2 +1500 S_1^3 S_2^3 +\ldots)\cr
d=8: &\qquad {1\over 40642560}(4 S_1^8 + 63 S_1^7 S_2 +441 S_2^6 S_1^2 + 441 S_1^5 S_2^3 +
30870 S_1^4 S_2^4 +\ldots)\ . }
$$
Both calculations are in perfect agreement.

\subsec{The genus one B-model amplitude}

According to \bcovI \bcovII\ and taking
the simplification in the local case \kz\ into account we expect the holomorphic
$\bar s_i\rightarrow 0$ limit of the topological amplitude to be
$$
F^{(1)}=\log \left(\det^{1\over 2} \left(\partial x_i\over \partial s_j\right)
\Delta(x_1,x_2)^{-{1\over 12}} \prod_{i=1}^2 x_i^{b_i}\right)\ ,
$$
where the conifold discriminant is given by $\Delta=(16 -16 x_2^2+ 8 x_1 x_2^2+x_1^2 x_2^4)$ in the $x$
coordinates. The exponent $-{1\over 12}$ at the conifold is universal and $b_1={1\over 3}$, $b_2=0$.
Note that the rescaling of the string coupling does not affect this comparison with the
expression from the matrix model. Expanding in the matrix model flat coordinates $(S_1,S_2)$ and get
$$
F^{(1)}=-{1\over 12}( \log S_1 +\log S_2) + \sum_{m,n} c^{(1)}_{m,n} S_1^m S_2^n.
$$
Again the $c^{(1)}_{m,n}$ are only non-zero for $n+m\in 2 \ZZ$ and symmetric in $m,n$.
The first few degrees are given by
$$
\eqalign{
d=2: &\qquad -{1\over 288}(S_1^2-6 S_1 S_2 + S_2^2)\cr
d=4: &\qquad {1\over 69120}(2 S_1^4+105 S_1^3 S_2-90 S_1^2 S_2^2+105 S_1 S_2^3+2 S_2^4)\cr
d=6: &\qquad -{1\over 17418240}
(8 S_1^6 - 189 S_1^5 S_2 + 7560 S_1^4 S_2^2 - 630 S_1^3 S_2^3 +\ldots)\cr
d=8: &\qquad {1\over 1857945600}(16 S_1^8 + 435 S_1^7 S_2 - 27195 S_2^6 S_1^2 + 196770 S_1^5 S_2^3 + 222600 S_1^4 S_2^4 +\ldots)\ .}
$$
in perfect agreement with the matrix model calculation \resone.

\subsec{The higher genus topological B-model amplitudes at the orbifold point}

The key problem in deriving higher genus results in the B-model with multi dimensional 
moduli space is to find the propagators of the topological B-model. Due to the 
technical nature of the problem we relegate the derivation of the propagators 
in the Appendix B. 

Equipped with $F^{(0)}$, $F^{(1)}$ and the propagator (B.7) $S:=S^{22}$ we can readily calculate 
$F^{(2)}$. Since we assured the same singular behavior of the propagator, 
the ambiguity  at genus $2$ has not to be determined again, but can be taken after 
suitable coordinate transformation from the calculation of the $F^{(2)}$ at the 
large complex structure.  
      
$$\eqalign{
F^{(2)}=& -{1\over 8} S_2^2F^{(0)}_{, 4} + 
{1\over 2}  S_2F^{(1)}_{, 2} + 
{5\over 24} S_2^3(F^{(0)}_{, 3})^2  - 
{1\over 2} S_2^2F^{(1)}_{, 1}F^{(0)}_{, 3} + 
{1\over 2} S_2(F^{(1)}_{, 1})^2+f^{(2)}\cr 
=&-{1\over 240}\left({1\over S^2_1}+ {1\over S^2_2}\right)+ \sum_{m,n} c^{(2)}_{m,n} S_1^m S_2^n}$$
The $c^{(2)}_{m,n}$ are only non-zero for $n+m\in 2 \ZZ$ and symmetric in $m,n$  
$$
\eqalign{
d=2: &\qquad -{1\over 57600}(S_1^2+60 S_1 S_2 + S_2^2)\cr
d=4: &\qquad {1\over 1451520}(S_1^4+126 S_1^3 S_2+ 378 S_1^2 S_2^2+126 S_1 S_2^3+S_2^4)\cr
d=6: &\qquad -{1\over 2654208000}
(64 S_1^6 - 38385 S_1^5 S_2 + 334575 S_1^4 S_2^2 + 124500 S_1^3 S_2^3 +\ldots)\cr
d=8: &\qquad {1\over  81749606400 }(64 S_1^8 + 68343 S_1^7 S_2 - 2224299 S_2^6 S_1^2 + 7547001 S_1^5 S_2^3 
+ 27188870 S_1^4 S_2^4 +\ldots)\ .}
$$
The $d=2$ term and the terms involving only one 
$S_i$ are again in perfect agreement with the matrix 
model \restwo\fghsimple.  

The iteration in the genus is in principle no problem in the B-model, however one has to fix the 
holomorphic ambiguity at each genus, which we pushed only up to genus 3.
      
$$\eqalign{
F^{(3)}
=&
S_2F^{(2)}_{, 1}F^{(1)}_{, 1}
-{1\over 2}S_2^2F^{(2)}_{, 1}F^{(0)}_{, 3} 
+ {1\over 2} S_2F^{(2)}_{, 2}
+ {1\over 6}S_2^3(F^{(1)}_{, 1})^3 F^{(0)}_{, 3}
- {1\over 2}S_2^2F^{(1)}_{, 2}(F^{(1)}_{, 1})^2 \cr &
- {1\over 2}S_2^4(F^{(1)}_{, 1})^2(F^{(0)}_{,3})^2 
+ {1\over 4}S_2^3(F^{(1)}_{, 1})^2F^{(0)}_{, 4}
+ S_2^3F^{(1)}_{, 2}F^{(1)}_{, 1}F^{(0)}_{, 3}
- {1\over 2}S_2^2F^{(1)}_{, 3}F^{(1)}_{, 1} \cr &
- {1\over 4}S_2^2(F^{(1)}_{, 2})^2
+ {5\over 8}S_2^5F^{(1)}_{, 1}(F^{(0)}_{, 3})^3
- {2\over 3}S_2^4F^{(1)}_{, 1}F^{(0)}_{, 4}F^{(0)}_{, 3}
- {5\over 8}S_2^4F^{(1)}_{, 2}(F^{(0)}_{, 3})^2 \cr &
+{1\over 4}S_2^3F^{(1)}_{, 2}F^{(0)}_{, 4}
+{5\over 12} S_2^3F^{(1)}_{, 3}F^{(0)}_{, 3}
+ {1\over 8}S_2^3F^{(0)}_{, 5}F^{(1)}_{, 1} 
- {1\over 8} S_2^2F^{(1)}_{, 4} 
- {7\over 48} S_2^4F^{(0)}_{, 5}F^{(0)}_{, 3} \cr &
+ {25 \over 48}S_2^5F^{(0)}_{, 4}(F^{(0)}_{, 3})^2 
-{5\over 16}S_2^6 (F^{(0)}_{, 3})^4   
- {1\over 12} S_2^4(F^{(0)}_{, 4})^2
+{1\over 48} S_2^3F^{(0)}_{, 6} 
+f^{(3)}
\cr 
=&-{1\over 1008}\left({1\over S^2_1}+ {1\over S^2_2}\right)+ \sum_{m,n} c^{(2)}_{m,n} S_1^m S_2^n\ . }
$$
The first few coefficients are
$$
\eqalign{
d=4: &\qquad {1\over 557383680}(16 S_1^4-345 S_1^3 S_2+ 58500 S_1^2 S_2^2-345 S_1 S_2^3+16 S_2^4)\cr
d=6: &\qquad -{1\over 36787322880}(64 S_1^6 - 325116 S_1^5 S_2 + 1461735 S_1^4 S_2^2 - 2198130 S_1^3 S_2^3 +\ldots)\ . }
$$
These results are predictions for the matrix model. 

\newsec{Some generalizations} 
\subsec{Adding matter}

The considerations in the preceding sections 
can be easily generalized by adding matter fields.
In terms of Chern-Simons theory this has been discussed in \refs{\amv,\dfg}.
In this section we consider this in the mirror B-model
language. We will show that all the amplitudes computed in
\refs{\amv,\dfg} are matrix model amplitudes. This includes 
invariants for torus knots and links in the classes of three- manifolds $M$ 
considered in this paper.

For definiteness, consider the B-model geometry corresponding
to
\eqn\four{xz = (e^{u}-1)(e^{u-t_1}-1)(e^{v}-1)(e^{v-t_2}-1),}
which is a mirror of the A-model geometry studied in section $7.5$ of \amv.
There are four holomorphic $\IP^1$'s corresponding to four points with
$u=0, t_1$ and $v=0,t_2$. We can consider
wrapping some numbers $N_i$ D-branes on the $i$-th $\IP^1$.
\ifig\fourspheres{The figure depicts the four isolated $\IP^1$'s in the
Calabi-Yau \four .}
{\epsfxsize 4.0truein\epsfbox{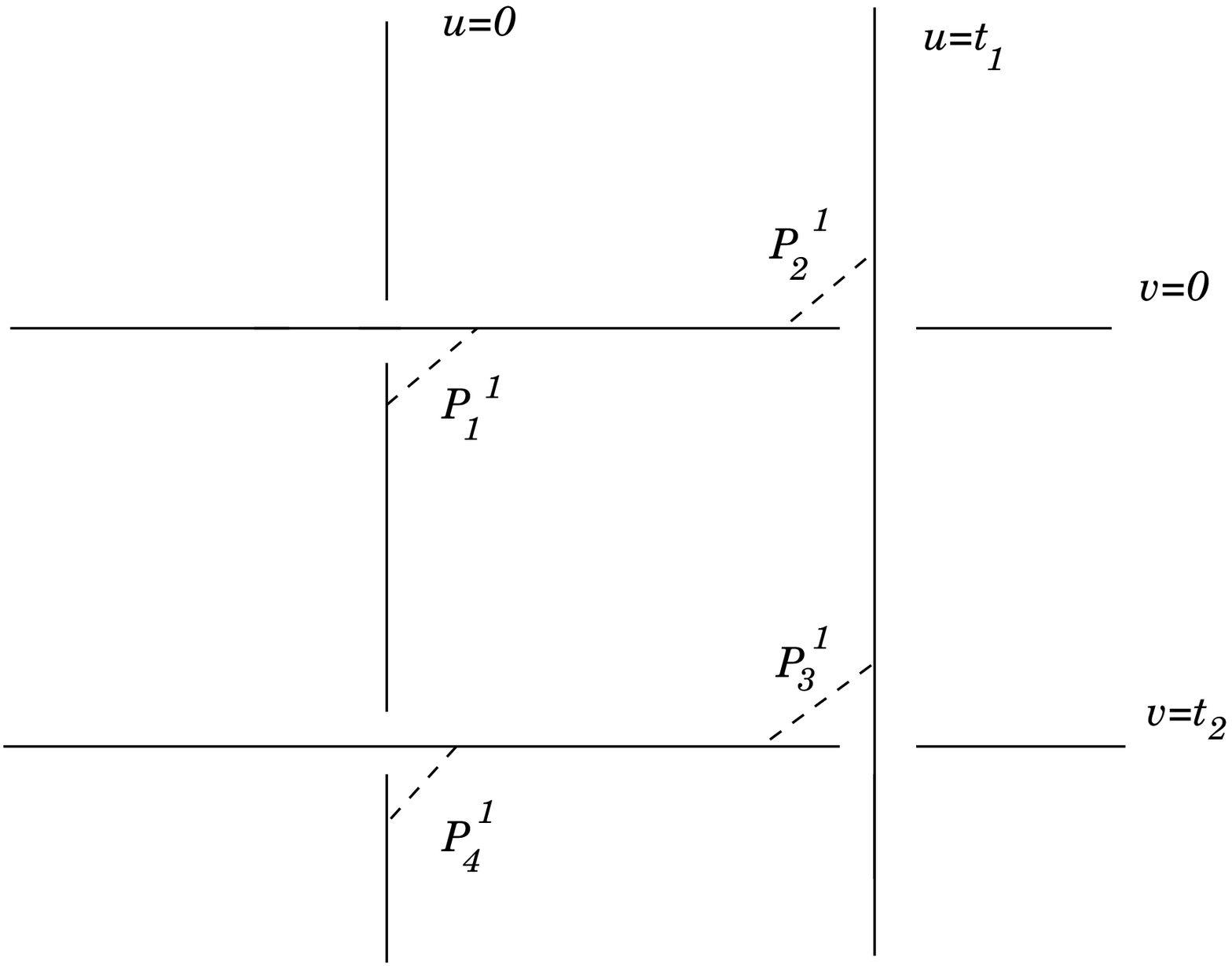}} 
Consider the partition function of the modes corresponding to
the $\IP^1$ at $u=0=v$. This is given by 
$$Z(S)  =\langle 0_v|S|0_v\rangle =\langle 0_u|0_v\rangle,$$
since $u$ and $v$ are related by
$S$ operator given in \slt. 
Alternatively,
the wave function $|0_u\rangle$ is obtained from 
$|0_v\rangle$ by simply exchanging $u$ and $v$, as this is what $S$ does,
up a constant. We have that
\eqn\roundsa{Z = \frac{1}{|{\cal W}|}
\int \prod_i 
\frac{d u_i d v_i}{(2\pi g_s)^{\frac{1}{2}}} \Delta_H(u)\Delta_H(v)\, e^{\sum_i  u_i v_i/g_s},}
where we used that $u$ and $v$ are canonically conjugate, so
$\langle u | v \rangle = e^{\sum_i\,  u_i v_i /g_s},$
and furthermore $|{\cal W}| = N!$ is the order of the Weyl group of $U(N)$
where $N=N_1$ is the number of wrapped D-branes.
It is easy to see that the partition functions of the modes living on
the other $\IP^1$'s in \four\ 
coincide with \roundsa\ with appropriate values of $N$.

This corresponds to a matrix model given by
\eqn\roundsb{Z = \frac{1}{{\rm vol}(U(N))}\int_{[u,v]=0} 
{{\hat d}u {\hat d }v \over (2\pi g_s)^{N \over 2}}  \, 
e^{{\rm Tr}\,  u v/g_s},}
where the integral is over commuting Hermitian matrices $u$ and $v$.
The measure in the path integral is defined as follows.
Consider the space of unitary matrices $U, V$. 
where $U=e^{u}$ and $V=e^{v}$. Since $u,v$ are canonically conjugate
it is natural to consider the symplectic form
\eqn\sym{\omega = {\rm Tr} \, U^{-1} dU\wedge V^{-1} dV\, ,}
in terms of the left $U(N)$ invariant line elements $U^{-1} dU$
and $V^{-1} dV$. The symplectic form gives rise to the 
volume element on the phase space $\omega^{d}/d!$ where $d=N^2$ is
the dimension of $U(N)$. The measure in the path integral \roundsb\
is induced from this by restricting to the space of commuting matrices $U$ and $V$.
Namely, $u$ and $v$ commute, there exists a unitary matrix 
$\Omega$ that diagonalizes both
$U$ and $V$, i.e. $\Omega U \Omega^{-1}={\rm diag}(u_i)$ and
$\Omega V \Omega^{-1}={\rm diag}(v_i)$. The volume
of the phase space is obtained by writing \sym\ in terms
of $\Omega$, $u_i$ and $v_i$.
Integrating over $\Omega$ to reduce the path integral to integral
over the eigenvalues recovers \roundsa. 
This is akin to the matrix models studied in 
\normal\ based on Hermitian matrices. In the following,
the measure on the phase space of pairs of conjugate, $U(N)$ Lie-algebra valued
variables, $u$, $v$ will be denoted by $\hat{d}u\hat{d}v$. In particular,
it should be understood that $u$ and $v$ commute.
The fact that $u$ and $v$ are commuting matrices in \roundsb\ is
natural as the $A_{\bar z}$ equation of motion
implies that the matrix
$[u(z),v(z)]_{ij}$
vanishes\foot{This is mirror to the vanishing of $F$ in \cst .}, 
and we have localized to zero modes.

Because there is more than one stack of B-branes,
there are additional open string sectors with the two ends of the string
on the D-branes wrapping the different $\IP^1$'s.
By the same arguments as in \amv, the only modes that contribute to the
B-model amplitudes correspond to the strings stretching between
$\IP^1_i$ and $\IP^1_{i+1}$ in the \fourspheres .

\ifig\strings{
The figure depicts the lift
in the full geometry of the line passing through
the north and the south pole of the $\IP^1_1$ and $\IP^1_2$. 
This is the $u$-cylinder intersected at
$u=0$ and $u_1$ by the two $\IP^1$'s. Consequently there is
a family of $BPS$ strings connecting the two $\IP^1$'s and
winding around the cylinder. The strings are labeled by their winding number.
}
{\epsfxsize 4.0truein\epsfbox{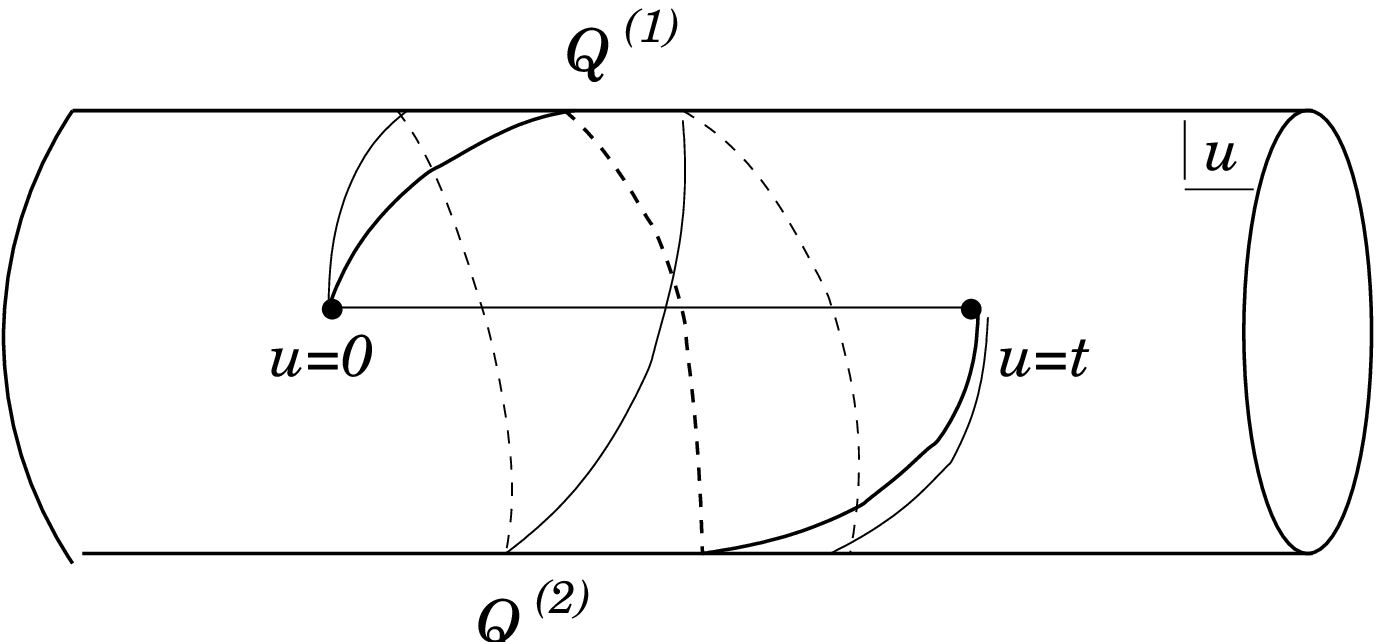}}

Consider for example the strings stretching between $u=0$ and $u=t_1$ 
on the $u$-cylinder and connecting $\IP^1_1$ and $\IP^1_2$.
There are different topological sectors of these strings -- corresponding
to how many times the string winds around the ${\bf S}^1$, see \strings.
From each sector we get one physical scalar in the 
bifundamental representation. Moreover, each of the strings 
is minimally coupled
to the gauge-fields on the spheres it ends on.
Thus, the matter part of the action is 
$$ S(Q_{12},u) = \sum_{n}\; {\rm Tr}\, Q^{(n)}_{12} \;\Bigl((2 \pi i n+t)\, 
1_1 \otimes 1_2 +\,u_1 \otimes 1_2 -\,u_1\otimes 1_2\Bigr)\;Q^{(n)}_{21},$$
where $[Q_{12}^{(n)}]^{\dagger} = Q_{21}^{(n)},$
and $u_1$ and $u_2$ are matrices corresponding to the positions of the first
and the second stack of D-branes on the $u$-cylinder. 
Recall that there is a relative shift by parameter $t$ between them that
contributes to the mass.
The contribution of this to the path integral 
${\cal O}_{12} = \int \prod_n\, dQ^{(n)}_{12}\,\exp(S(Q_{12},u)/g_s)\;$ is trivial to evaluate, 
as in \ovknot, giving
\eqn\ove{\eqalign{
{\cal O}_{12}=&
\exp\Bigl\{\,\sum_{n=1}^{\infty}\; 
\frac{{\rm e}^{-nt}}{n} \;{\rm Tr} U_1^{n}\; {\rm Tr}U_2^{-n}\Bigr\}\cr
=& \sum_R \;e^{-l_R t}\;{\rm Tr}_R U_1 \,{\rm Tr}_R U_2, }}
where $U_{1,2} = e^{u_{1,2}}$ and $l_R$ is the number of boxes in representation $R$.
In writing \ove\ we used the regularization
$\sum_{n} \log (2 \pi i n+ x) = \log\sinh (x) +\,const$ of the one 
loop path integral.

Note that mirror symmetry transforms the tower of modes above
to a single string ground state propagating on a circle, 
where as above $u_{1,2}$ get related to Wilson lines on the mirror ${\bf S}^3$'s 
as in \amv :
$S(Q_{12},A)=  \oint_{\gamma}\;{ \rm Tr} \; Q_{12}\;\Bigl((d+ t) 1_1 \otimes 1_2 -A_1 \otimes 1_2+ 1_1\otimes A_2\Bigr)\;Q_{21}$.
This is the expected action of T-duality on D-branes. 
The operators $U_{1,2}$ 
are now interpreted as Wilson loop operators in Chern-Simons
theory.

The matrix model allows one to very simply calculate expectation 
values of Wilson loop operators. 
Consider for example evaluating the Chern-Simons path integral 
on ${\bf S}^3$ in
the presence of a Hopf link. This can be obtained from
gluing two solid 2-tori by an $S$ transformation, and
in the presence of Wilson loops in representations $R$ and $R'$ 
on the one-cycles that cannot shrink, {\it i.e.} we 
are interested in evaluating 
$\langle R| S| R' \rangle = \langle R_{v}\, |\,R'_u\rangle$. 
In the light of the discussion above,
the path integral on the solid two-torus with the Wilson loop
is mirror to computing
$$|R_{v}\rangle=\frac{1}{|{\cal W}|^{\frac{1}{2}}}\int \prod_i du_i {\rm Tr}_R U \Delta_H(u) |u \rangle ,$$
so that $\langle R| S| R' \rangle$ is
\eqn\hopf{\langle R|  S| R' \rangle=\frac{1}{{\rm vol}(U(N))} \,\int
\,{\hat{d}u \hat{d}v\over (2\pi g_s)^{\frac{N}{2}}} {\rm Tr}_R U 
e^{{\rm Tr} \, u v /g_s} {\rm Tr}_{R'} V^{-1}.}
It is easy to see that this agrees with the expression \matsl\ --
as above it can be evaluated simply by using the Weyl character formula,
for $U(N)$
$${\rm Tr}_R U = \frac{\sum_{w \in {\cal W}} \epsilon (w)
{\rm e}^{iw (\alpha) \cdot u}}
{\sum_{w \in {\cal W}} \epsilon(w) {\rm e}^{i w (\rho) \cdot u}},$$
where $\alpha$ is the highest weight vector of the 
representation $R$ of $U(N)$ shifted by $\rho$.
As a check, note that
\eqn\hopf{\eqalign{
\langle R|  S| R' \rangle = &
\frac{1}{|W|} \Bigl( {g_s \over {2 \pi}} \Bigr)^{N\over 2}\sum_{w,w' \in {\cal W}} \epsilon (w w')
{\rm e}^{-g_s w (\alpha) \cdot w'(\beta)}\cr
=&\Bigl({g_s  \over  2 \pi}\Bigr)^{N\over 2}\;
{\sum_{w \in {\cal W}} \epsilon (w){\rm e}^{-g_s w (\alpha) \cdot \beta}}}.}
Note that this exactly agrees with \matsl .

To summarize, we have a matrix model expression for
the B-branes in the geometry \four, given by
\eqn\fourp{\eqalign{
Z_{tot} = \frac{1}{\prod_{i=1}^{4}{\rm vol}(U(N_i))}\;
\sum_{R_1,\ldots,R_4}&\; \int_{[u_i,v_i]=0} \prod_{i=1}^{4}\frac{\hat{d}u_i
\hat{d}v_i}{(2 \pi g_s )^{N_i\over 2}}
\;e^{-\sum_{i=1}^{4}{\rm Tr} \, u_i v_i /g_s} \cr
\times & e^{-l_1 t_1} \,{\rm Tr}_{R_1} U^{-1}_2  \,{\rm Tr}_{R_2}V_2 
 e^{-l_2
t_2}\, {\rm Tr}_{R_2}V_3^{-1}\, {\rm Tr}_{R_3}U_3 \cr
\times & e^{-l_3 t_1} \, {\rm Tr}_{R_3} U_4^{-1}\, {\rm Tr}_{R_4}V_4 \,e^{-l_4 t_2}
{\rm Tr}_{R_4} V^{-1}_{1}\,{\rm Tr}_{R_1} U_1}}
where the expectation values of the Hopf link operators are computed
by matrix integrals \hopf . The minus sign in the exponent corresponds to
the fact that the gluing operator is $S^{-1}$ \amv .
\ifig\twotran{The figure depicts the geometric transition
of the open string geometry in figure \fourspheres . The geometric
transition is a large $N$ duality, and the matrix model computes
amplitudes of the A-model version of geometry on the left (a toric $\IB_5$) 
to all genera.}
{\epsfxsize 4.0truein\epsfbox{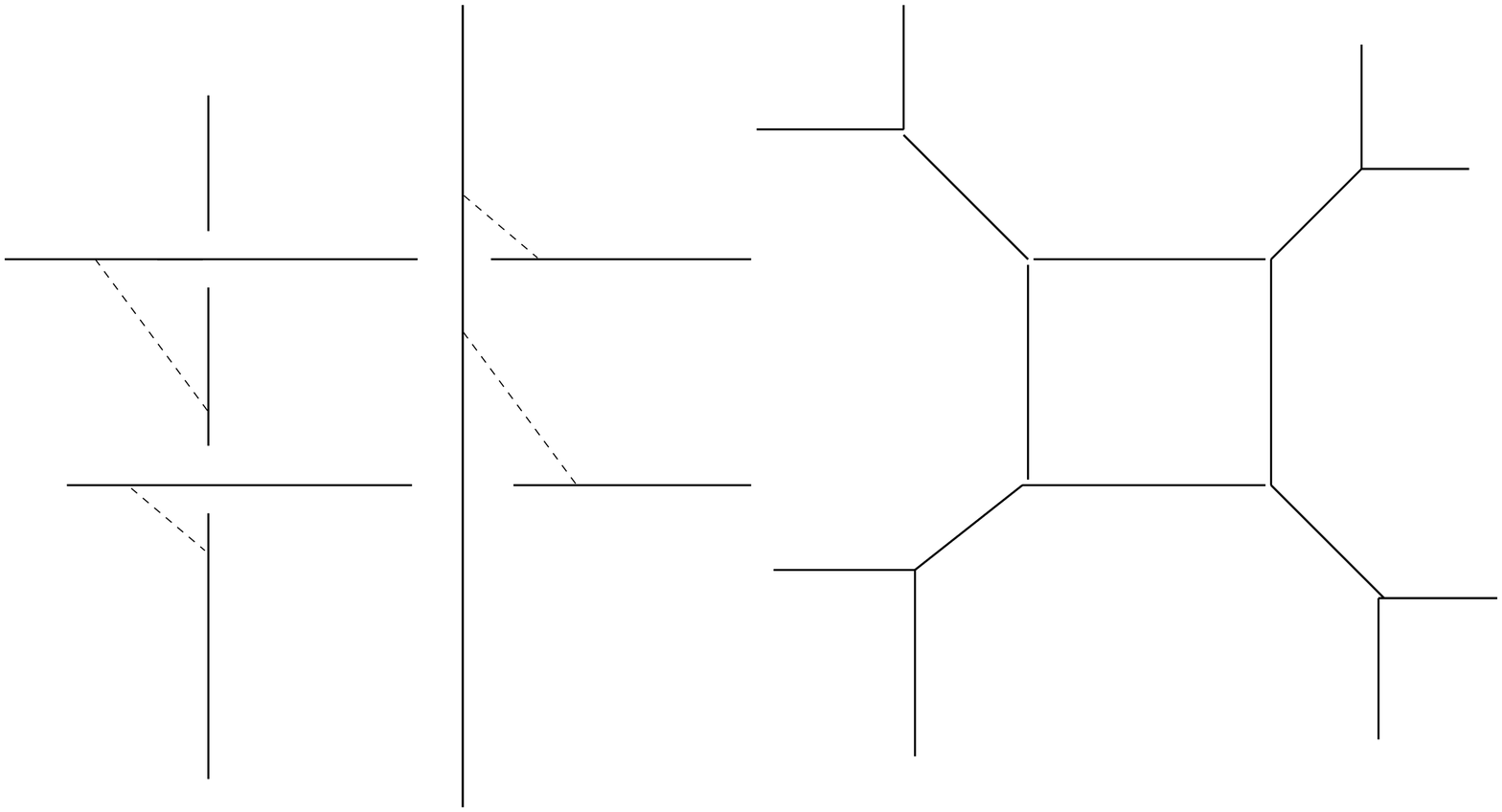}} 
Moreover, the large $N$ dual of this is a $mirror$ of 
toric Calabi-Yau geometry without
any D-branes, corresponding to a $O(-K) \rightarrow \IP^1 \times \IP^1$ 
blown up at four points, which is a non-generic Del-Pezzo surface $\IB_5$, 
see \twotran . 
In \amv\ it was shown that Chern-Simons theory on the A-model mirror of
geometries in \four\ computes the topological A-model amplitudes
in $\IB_5$ to all genera. What we have shown here is that
all genus $\IB_5$ amplitudes are really computed by a matrix model!

The above Hopf-link computation 
is also easy to generalize to more general $(m,n)$ torus knots,
where the corresponding operator is 
$$|R;m,n\rangle = {\rm Tr}_R e^{mu + n v}|0_v\rangle.$$
where we have picked one particular ordering of operators.
Different orderings of the operators differ from this by overall phases.

\subsec{More general geometries}

The considerations above can be generalized to arbitrary
backgrounds of the form
$$xz=\prod_i P_i(u,v).$$
We have a collection of $\IP^1$'s where curves
$$P_i = 0 = P_j$$
intersect.
In general there are also matter multiplets corresponding to
strings stretching between the $\IP^1$'s whose poles lie on the same curve,
and we get a quiver theory.
On the nodes of the quiver we get a matrix model,
$$ Z_{ij} = \langle0_{P_i}\;|\; 0_{P_j}\rangle.$$
For example, if $P_i=0$ and $P_{j} =0$ are given respectively  by
\eqn\solve{v = W_i'(u),\quad u = W_j'(v),}
then 
\eqn\genb{\eqalign{ Z_{ij} = &
\;\langle 0_v\,|\;{\rm e}^{W_i(u)/g_s} \;{\rm e}^{W_j(v)/g_s}\;|\,0_u\rangle\cr
=& \frac{1}{{\rm vol}(U(N))}
\int \,{\hat{d} u\; \hat{d} v \over (2 \pi g_s )^{N\over 2}}\; 
\exp\Bigl\{ \frac{1}{g_s} \Bigl({\rm Tr}\,W_i(u)-{\rm Tr}\, uv + 
{\rm Tr}\, W_j(v)\Bigr) \Bigr\} ,}}
This corresponds to replacing $T^*M$ by 
more general geometries which approximate this in the immediate
neighborhood of $M$.
\ifig\old{An example of more general geometries.}
{\epsfxsize 3.0truein\epsfbox{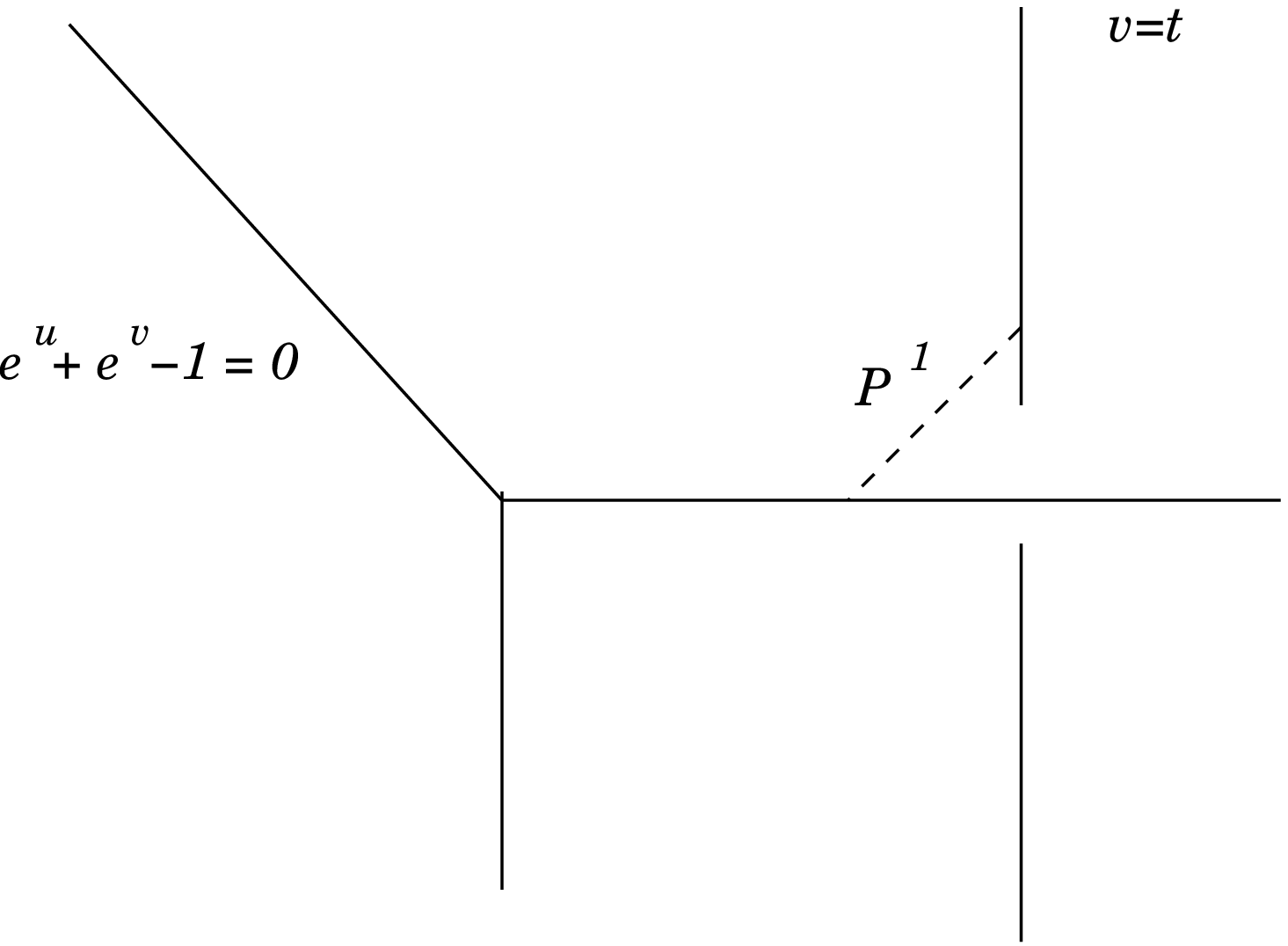}} 
As an example, let us consider the B-model geometry studied previously in
\AV,
corresponding to a blowup of 
$$xz = (e^{u} + e^{v} -1) ( e^{v-t} -1),$$
we have $W_1(v) = - \sum_{n} \frac{1}{n^2 }e^{nv}$ and
$W_2(u) = t u$.
Note that the gluing operator is
related to the superpotential of the theory.
Namely, 
the superpotential for a B-brane in this geometry was computed in \AV,
where it was found that
$$W(u)=  \int_B \Omega = \int (v_2(u)- v_1(u)) du,$$
where
$$v_1= log(1-e^{u}),\quad \quad v_2= t,$$ 
are the equations of the corresponding
Riemann surfaces. 
In calculating the superpotential, we kept $v$
fixed over the whole $\IP^1$. Note
that on the Riemman surface, if we put $v_1(u) = \frac{d}{du}W_1^{D}(u)$,
we have that $W^{D}_1(u)$ and $W_1(v)$ are Legendre transforms 
$$W^{D}_1(u) = uv + W_1(v).$$

\subsec{Framing dependence}

In the formalism we have been developing there is a subtlety related to
framing dependence. Note that there is more than one operator
having the property that it conjugates $v$ to $pu+qv$.
For example, given an operator $U_{(p,q)}$ that conjugates $v$ to $pu+qv$,
operator $U'_{(p,q)}$ 
$$U'_{(p,q)}  = U_{(p,q)}e^{-m v^2/2 g_s},$$
has the same property, for any value of $m$.
This corresponds geometrically to a ${\rm SL}(2,\IZ)$ transformation
that leaves the shrinking one-cycle of the boundary $T^2$
invariant $T: \;b \rightarrow b,$
but affects the finite cycle $a\rightarrow a + m b$. The 
resulting ambiguity is related to
the choice of framing in Chern-Simons theory,
and affects the vacuum expectation value by
an overall phase that one can readily calculate, so presents
no loss of predictability.

In fact, we can derive the known framing dependence by using the matrix
model representation. Consider the solid torus with a Wilson line in 
representation $R$.
Changing framing affects the path integral 
$|R_v\rangle = \sum_{\omega \in {\cal W}}\epsilon(\omega) \delta(v' +ig_s \omega(\alpha))$ as
\eqn\fram{\eqalign{|R_v\rangle &\ \rightarrow \  
e^{\frac{-m v^2}{2 g_s}}|R_v\rangle\cr
&\ = \  \sum_{w \in {\cal W}}\epsilon(w) 
\int dv' e^{\frac{m (v')^2}{2 g_s}}|v' \rangle 
 \delta(v' +i g_s w(\alpha))\cr  
&\ = \ e^{\frac{m g_s {\alpha \cdot \alpha}}{2}}|R_v\rangle.}}
Recall that $\alpha$ is the highest weight vector $\lambda$ 
of representation $R$
shifted by the Weyl vector, i.e. $\alpha = \Lambda+\rho.$
From this we see that
$\alpha \cdot \alpha = C_R + \rho \cdot \rho,$
where $C_R$ is the quadratic Casimir of the representation $R$.
Note that $\rho \cdot \rho=\frac{N(N^2-1)}{12} $. Therefore, the state
$|R_v\rangle$ gets multiplied by a relative phase
$$
\exp (2 \pi i m h_R)
$$
where $h_R= \frac{C_R}{2 (k+N)}$ is the conformal weight of the primary 
field in representation $R$ of the corresponding WZW model. The above
result is the well-known framing dependence of Wilson lines in CS theory. 
The remaining phase, $\exp(m g_s \rho^2/2)$, corresponds in fact 
to a change in the framing of the three-manifold.
Namely, $ g_s \rho^2/ 2 = -2 \pi i c_{U(N)}/24$ up to a constant $2\pi
N^2/24$.

\newsec{Relation to ${\cal N}=1$ theories}

Consider IIA theory compactified on $T^* {\bf S}^3$ with $N$ 
D6 branes wrapping ${\bf S}^3$. At low energies, the theory in four dimensions 
reduces to ${\cal N}=1$ super Yang-Mills, and as it was shown in \bcovII\sts\ the open string amplitudes $F_{g,h}$ 
lead to superpotential terms in the effective four-dimensional theory of the form
$$
\int d^2\theta F_{g,h} {\cal W}^{2g}[Nh S^{h-1}],$$
where ${\cal W}_{\alpha \beta}$ is an ${\cal N}=1$ multiplet whose 
bottom component is the 
self-dual part of the graviphoton, and $S={\rm Tr}W_{\alpha}W^{\alpha}$ is
the gluino superfield. Notice that the derivative with respect to $S$ of
the prepotential $F_0 (S)=\sum_h F_{0,h} S^h$ gives the superpotential of the 
${\cal N}=1$ theory. 

The small $S$ behavior of the superpotential is captured by the leading 
piece of $F_0(S)$, in other words, by the behavior of the prepotential near 
the conifold point 
$$
F_0(S) = {1\over 2} S^2 \log\, S.
$$  
As shown in \sts, this leads to the Veneziano-Yankielowicz gluino 
superpotential. The full prepotential is given by 
$$
F_0(S) = {1\over 2} S^2 \log\, S + \sum_{h=4}^\infty {B_{h-2} \over (h-2)h!} S^h,
$$
and leads to a superpotential that can be written (in string units) as \sts\
\eqn\effw{
W=\sum_{n \in \IZ} (S+ 2\pi in)\log (S+ 2\pi i n)^{-N}.}
This can be interpreted in terms of infinitely many species of domain walls 
labeled by $n$ \sts. With the results of this paper we can give another
interpretation of \effw. According to the general result of \dva\dvthree, 
the effective superpotential
of an ${\cal N}=1$ supersymmetric gauge theory can be computed by a matrix
model whose potential is the tree level superpotential of the gauge
theory. On the other hand, we have seen 
that there is a Hermitian matrix model describing Chern-Simons theory on the three-sphere, given by \herm. This means that \effw, which includes infinitely 
many domain walls, can be interpreted as the effective superpotential of an 
${\cal N}=2$ theory whose tree-level superpotential is
\eqn\super{
{1 \over 2} {\rm Tr} \Phi^2 + S  \sum_{k=0}^{\infty}{B_{2k} \over (2k) 
(2k)!} \sum_{s=0}^{2k} (-1)^{s-1} {2k \choose s} {1 \over N} {\rm Tr} 
\Phi^s {\rm Tr}\Phi^{2k-s}.
}
Here, $\Phi$ is the ${\cal N}=1$ chiral superfield 
in the adjoint representation which is part of the 
${\cal N}=2$ vector multiplet, and we have used that $g_s \rightarrow S/N$ 
\refs{\sts,\dva,\dvthree}. Notice that this
superpotential contains multi-trace operators. These kinds of operators
have been recently considered in the context of the AdS/CFT correspondence,
see for example \refs{\silver,\multi}.

A similar argument can be applied to type IIA theory compactified on 
$T^*M_p$, with $N$ D6 branes wrapping $M_p$. Since we are orbifolding
with a $\IZ_p$ action, we have in general a quiver theory 
with $p$ nodes and gauge groups $U(N_1) \times \cdots U(N_p)$. Each of
these 
quiver theories ({\it i.e.} the different choices of $N_1, \cdots, N_p$) are in
one-to-one correspondence with the choices of flat connections in the
corresponding Chern-Simons theory. At leading order in the gluino 
superfields, this theory is just a direct product of $U(N_i)$ theories that
do not interact with each other, and the 
prepotential is just the sum of the
corresponding prepotentials for the different gauge
groups. However, as we have seen in this paper, the higher order
corrections mix the different gluino fields, and we can interpret the
resulting ${\cal N}=1$ superpotential as coming from a product of $p$ ${\cal
N}=2$ 
theories with gauge groups $U(N_1), \cdots, U(N_p)$, and with a tree level
superpotential that can be read from \generalp:
$$
{1 \over 2 }
\sum_{i=1}^p {\rm
Tr}\Phi_i^2 + {S \over p N} \sum_{i=1}^p V(\Phi_i) - {S \over p N}
\sum_{1\le i<j \le p} W(\Phi_i, \Phi_j),
$$
where $\Phi_i$ is the ${\cal N}=1$ chiral superfield in the adjoint of the 
$U(N_i)$ theory,
$V(\Phi)$ is given by the second term in \super, and $W(\Phi_i, \Phi_j)$ is 
given in \hermint.

\bigskip

\centerline{\bf Acknowledgements}
We would like to thank E. Diaconescu, R. Dijkgraaf, M. Douglas,
S. Garoufalidis, J. Gomis, R. Gopakumar, A. Grassi,
S. Gukov, G. Moore, B. Pioline, S. Sinha, and S. Theisen for 
useful discussions. The work of M.A., M.M. and C.V. is supported in part by 
NSF grants PHY-9802709 and DMS-0074329. A.K. is supported in part by the DFG 
grant KL-1070/2-1.

\appendix{A}{Computation of correlation functions in the Gaussian matrix model}
In order to compute the perturbation expansion of \efftwo, one has to evaluate 
correlation functions in the Gaussian matrix model. In this short appendix we 
review some basic techniques to do these computations. 
\ifig\vertices{Fatgraphs representing ${\rm Tr}M$, ${\rm Tr}M^2$ 
and ${\rm Tr}M^3$.}
{\epsfxsize 3.0truein\epsfbox{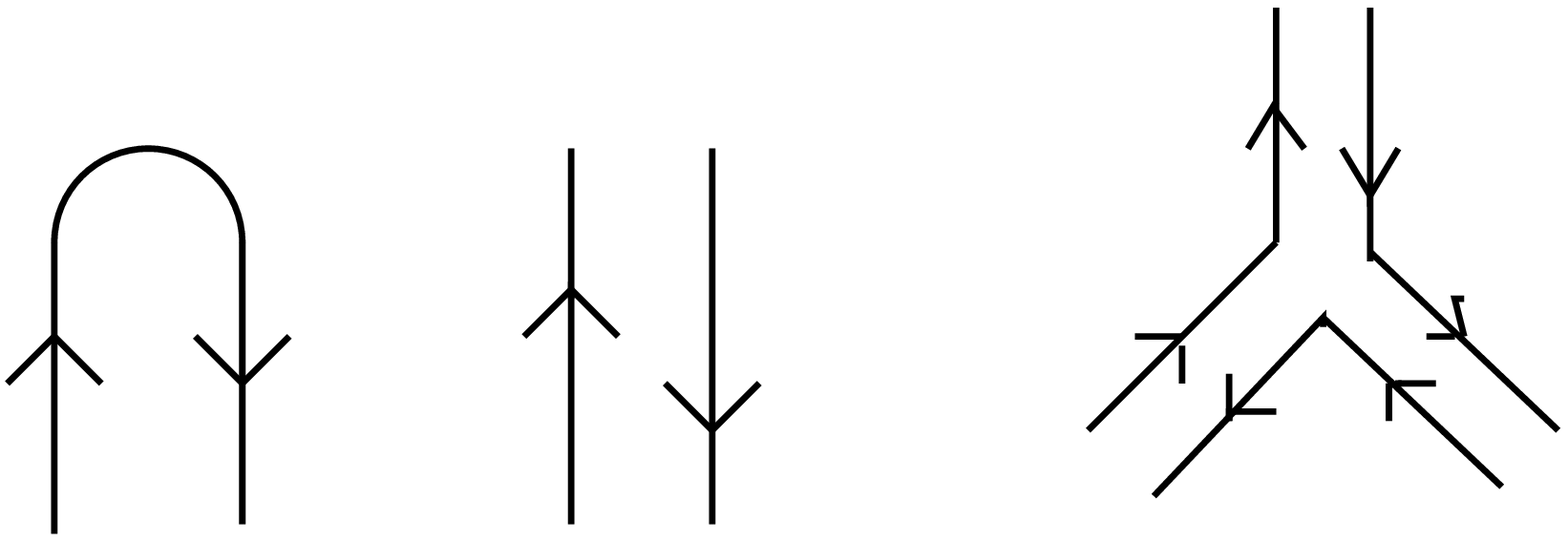}}

We want to evaluate normalized correlation functions of the form
\eqn\normvev{
\langle \prod_j ({\rm Tr} M^j)^{k_j} \rangle
= { \int dM {\rm e}^{ -{1 \over 2} {\rm Tr} M^2} 
\prod_j ({\rm Tr} M^j)^{k_j}  \over 
\int dM {\rm e}^{ -{1 \over 2} {\rm Tr} M^2}}, }
where $M$ is an $N \times N$ Hermitian matrix.
When the exponent of the Gaussian is given by 
$-{1 \over 2 \hat g_s} {\rm Tr}\, M^2$, the 
above correlation functions gets multiplied by $\hat g_s^\ell$, where 
$\ell=\sum_j j k_j$. Notice that the correlation function is different 
from zero only when $\ell$ is even. 

There are 
various ways to obtain the value of \normvev. A useful technique is to 
use the matrix 
version of Wick's theorem, or its graphic implementation in terms of fatgraphs (see 
\difr\ for a nice review). An insertion of $({\rm Tr}M^j)^{k_j}$ 
leads to $k_j$ $j$-vertices written in the double line notation, 
and the average \normvev\ is 
evaluated by performing all the contractions. The propagator is the 
usual double line propagator. Each resulting graph $\Gamma$ 
gives a power of $N^{\ell}$, where $\ell$ is the 
number of closed loops in $\Gamma$. 
Since we have insertions of ${\rm Tr} M$ as 
well, we have to consider a one-vertex given by 
a double line in which two of the ends have
been joined. The one, two and three-vertices in terms of fatgraphs are 
shown in \vertices. \ifig\mcube{The fatgraph contributing to $\langle {\rm Tr}M
{\rm Tr}M^3\rangle$.}
{\epsfxsize 1.75truein\epsfbox{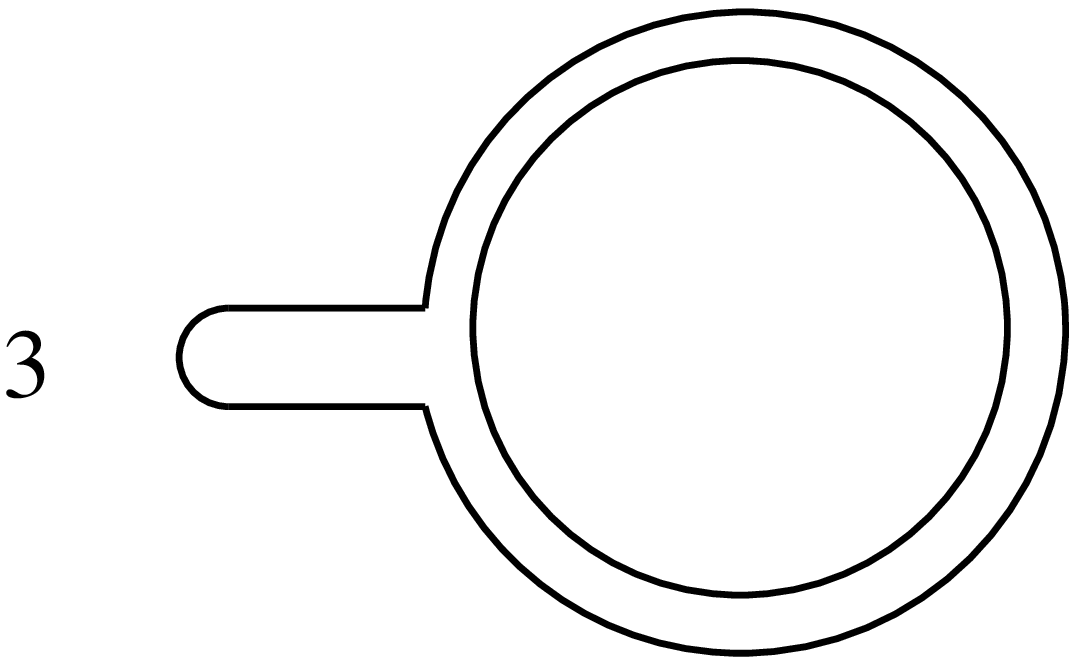}} 
As an example, consider the average 
\eqn\exam{
\langle \sigma_2 \rangle=\langle N {\rm Tr}M^4 -4 {\rm Tr}M {\rm Tr}M^3 
+ 3 ({\rm Tr}M^2)^2 \rangle.}
The evaluation of  $\langle  {\rm Tr}M^4 \rangle$ is standard \biz: we have
one planar diagram with weight
$2$  giving
$2N^3$, and one nonplanar diagram 
(with $g=1$) giving $N$. 
In the evaluation of  $\langle {\rm Tr}M
{\rm Tr}M^3\rangle$ we have three possible contractions between the
one-vertex and the three-vertex of \vertices, leading 
to a planar diagram with weight $3$, 
as shown in \mcube. Since there are two closed loops, the final result 
is $3N^2$. \ifig\msq{The fatgraphs contributing to 
$\langle ({\rm Tr}M^2)^2 \rangle$.}
{\epsfxsize 4.0truein\epsfbox{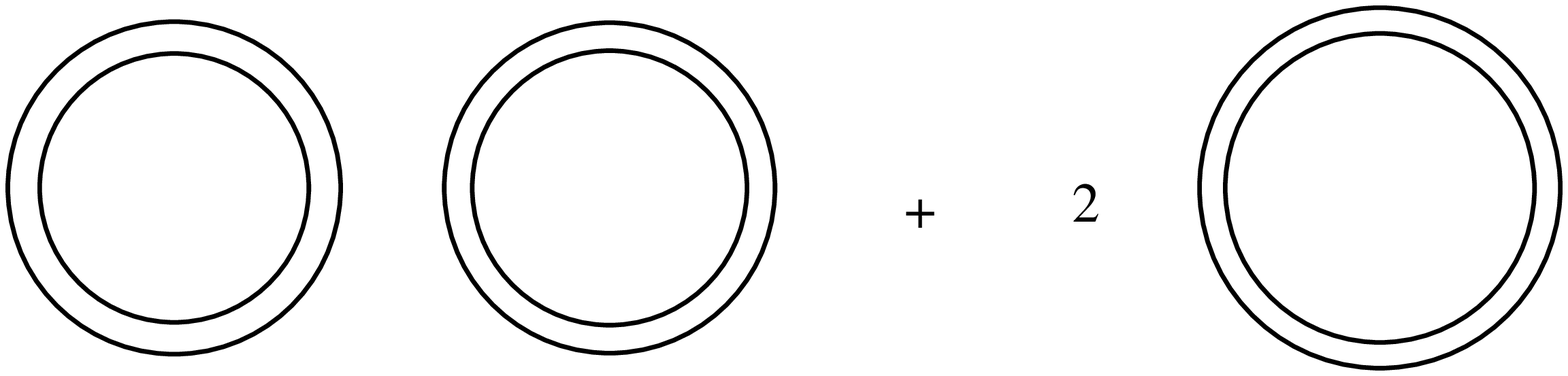}}
To evaluate $\langle ({\rm Tr}M^2)^2 \rangle$ we consider two
two-vertices. We can do self contractions, leading to one disconnected planar
diagram with four loops, or we can contract the two-vertices one to another 
in two ways, leading to a connected diagram with two loops, see \msq. We
find in total $N^4 + 2 N^2$. Putting everything together, we obtain:
$$
\langle \sigma_2 \rangle =5N^2(N^2-1).
$$

It turns out that one can write a general an explicit expression for the
average \normvev\ using results of Di Francesco and Itzykson \dfi. 
This goes as follows. By Frobenius formula, one can 
express the product of traces $\prod_j ({\rm Tr} M^j)^{k_j}$ as a linear 
combination of traces in irreducible representations $R$. To do that, one 
regards the vector $(k_1, k_2, \cdots)$ as a conjugacy class of the
symmetric group of $\ell=\sum_j j k_j$ elements. This conjugacy class, that 
we will denote by $C(\vec k)$, has $k_1$ cycles of length $1$, $k_2$ cycles
of length $2$, and so on. We then have,
\eqn\frob{
\prod_j ({\rm Tr} M^j)^{k_j} =\sum_R \chi_R (C(\vec k)) \, {\rm Tr}_R M,}
where the sum is over representations of the symmetric group of $\ell$
elements. These representations are associated to Young tableaux with
$\ell$ boxes, and we will denote the number of boxes in the $i$-th row of
the Young tableau by $l_i$, with $l_1 \ge l_2 \ge \cdots$. Define now 
the $\ell$ integers $f_i$ as follows
\eqn\fis{
f_i=l_i +\ell -i,\,\,\,\,\,\, i=1, \cdots, \ell.}
We will say that the Young tableau associated to $R$ 
is even if the number of odd $f_i$'s is
the same as the number of even $f_i$'s. Otherwise, we will say that it is
odd (remember that $\ell$ is even). One can show \dfi\ that the average of ${\rm Tr}_R M$ in the 
Gaussian matrix model vanishes if $R$ is an odd tableau, and for even 
tableaux one has the explicit formula:
\eqn\aver{
\langle {\rm Tr}_R M \rangle = (-1)^{A(A-1)\over 2} 
{\prod_{f \, {\rm odd}} f!! \prod_{f' \, {\rm even}} f'!! 
\over \prod_{f \, {\rm odd}, f'\, {\rm even}} (f-f')} d_R,}
where $A=\ell/2$. Here $d_R$ is the dimension of $R$ as an 
irreducible representation of
$U(N)$, and can be computed for example by using the hook formula. As 
an example of \aver, let us compute $\langle {\rm Tr} M 
{\rm Tr}M^3 \rangle$. To do that, one has to evaluate $\langle {\rm Tr}_R M 
\rangle$ for $R=\tableau{4}$, $\tableau{2 2}$ and $\tableau{1 1 1 1}$. 
All of these tableaux are even, and one finds:
\eqn\restab{
\eqalign{
 \langle {\rm Tr}_{\tableau{4}} M 
\rangle=& {1 \over 8} N (N+1)(N+2) (N+3),\cr  
\langle {\rm Tr}_{\tableau{2 2}} M 
\rangle=& {1 \over 4} N^2 (N^2-1), \cr
\langle {\rm Tr}_{\tableau{1 1 1 1}} M 
\rangle=& {1 \over 8} N (N-1)(N-2) (N-3).\cr}}
One then finds, by using Frobenius formula, 
\eqn\frobmmcube{
\langle {\rm Tr} M 
{\rm Tr}M^3 \rangle= \langle {\rm Tr}_{\tableau{4}} M 
\rangle - \langle {\rm Tr}_{\tableau{2 2}} M 
\rangle + \langle {\rm Tr}_{\tableau{1 1 1 1}} M 
\rangle  = 3 N^2,}
in agreement with the result that we obtained with fatgraphs. 

Although the result of \dfi\ explained above gives a general answer, 
in some cases there are more convenient expressions. For example, 
for $\langle {\rm Tr}\,M^{2j+2} \rangle$, Kostov and Mehta \km\ 
found the useful result:
\eqn\kmeq{
\langle {\rm Tr}\,M^{2j+2} \rangle= {(2j+2)! \over (j+1)! (j+2)!} 
P_{j+1}(N),}
where
\eqn\pj{
P_{m}(N)=\sum_{i=0}^{[m/2]}{ a_{mi} \over 4^i}N^{m+1-2i},}
and the coefficients $a_{mi}$ are defined by the recursion relation
\eqn\recur{
a_{m+1,i}=\sum_{k=2i-1}^m k(k+1) a_{k-1, i-1},}
and $a_{m0}=1$. One has for example $a_{m1}=(m+1)m (m-1)/3$, and so on. 
Notice that in the planar 
limit, the leading term of the 
average \kmeq\ is given by $N^{j+2}$ times the Catalan number 
$c_{j+1}=(2j+2)! /((j+1)! (j+2)!)$. Using \kmeq, one finds for example,
$$\eqalign{
\langle {\rm Tr}\, M^4 \rangle =& 2N^3 + N, \cr
\langle {\rm Tr}\, M^6 \rangle =& 5N^4 + 10 N^2, \cr
\langle {\rm Tr}\, M^8 \rangle =& 14 N^5 + 70 N^3 + 21 N. \cr}$$

Finally, another useful fact in the computation of \normvev\ is that 
averages of the form $\langle ({\rm Tr}M^2)^p\, {\cal O} \rangle$ can 
be evaluated 
by restoring appropriately the $g=1/\hat g_s$ dependence in the 
Gaussian. One easily finds that, if ${\cal O}$ is an 
operator of the form $\prod_j ({\rm Tr} M^j)^{k_j}$, with 
$\ell =\sum_k j k_j$, 
then 
$$
\langle ({\rm Tr}M^2)^p \, {\cal O} \rangle= 
\biggl( -2{d \over dg}\biggr)^p g^{-{\ell+ N^2\over 2}} 
\Big|_{g=1} \langle {\cal O} \rangle.$$

\appendix{B}{Derivation of the propagators in the B-model}

One main problem in the analysis of the B-model is the determination of the 
propagators $S^{ij}$ with the defining relation ${\bar \partial}_{\bar i}
S^{kl}={\bar C}^{kl}_{\bar i}$ \bcovII.
They are simply integrated w.r.t.
${\bar \partial}_{\bar \jmath}$ from the special geometry relation
\eqn\specialgeometry{R_{i\bar \jmath l}^k=G_{i\bar \jmath} \delta_l^k+G_{k
    \bar \jmath} \delta_i^k - C_{ilm} {\bar C}^{km}_{\bar \jmath},}
using the wellknown formulas in K\"ahler geometry
$R_{i\bar \jmath l}^k=-{\bar \partial}_{\bar \jmath} \Gamma^{k}_{il}$,
$G_{i\bar \jmath}=\partial_i{\bar \partial}_{\bar \jmath} K$ and
$\Gamma^{i}_{lm}=G^{i \bar k}\partial l G_{\bar k,m}$ to
\eqn\propeq{S^{ij} C_{jkl}=\delta_l^i\partial_k K+\delta^i_k\partial_l
  K+\Gamma^{i}_{kl}+f^{i}_{kl}. }
However there are two problems in actually solving for the $S^{ij}$.
The purely holomorphic terms $f^{i}_{kl}$ are ambiguous
integration constants. In the multi moduli case the ${1\over 2} n^2(n+1)$
equations overdetermine the ${1\over 2} n(n+1)$ $S^{ji}=S^{ji}$, $i\leq
j$ and the $f^{i}_{kl}$ can in general not be trivial. Secondly, since
the left hand side of \propeq\ is covariant, the $f^{i}_{kl}$ have to undo
the inhomogeneous transformation of Christoffel symbol as well as the shift
of the first two terms of the left hand side under K\"ahler transformations.

While for the instanton expansion we need the flat large complex
structure variables $t_i(z_k)$, we expect the $f^{i}_{kl}$ to be simple rational
functions involving the discriminant components in the $z_i$ variables,
because the $C_{ikl}$ have similar properties in these coordinates. We will
first solve the problem of the over determination of the $S^{ij}$ in
the $z$ coordinates and then transform the $S^{ij}$ as covariant tensors
to the $t$ variables, this determines the choice of the $f^{i}_{jk}$ in the
$t$ variables.

Let us discuss some non-compact Calabi-Yau manifolds first.
Here we have the simplification that in the holomorphic limit
the K\"ahler potential becomes a constant and furthermore there is a gauge
\kz\ in which the propagators ${\bar \partial}_{\bar \jmath} S^{j}:=
S^{j}_{\bar \jmath}$ and ${\bar \partial}_{\bar \jmath} S:= S_{\bar \jmath}$
vanish in that limit, which makes the topological amplitudes entirely
independent from quantities like the Euler number or the Chern classes,
which would have to be regularized in the non-compact case.

The simplest cases to consider are ${\cal O}(-2,-(n+2))\rightarrow F_n$.
We use the parameterization of the complex structure variables of \kkvI \ckyz ,
where the Picard-Fuchs equations and the genus zero and genus one
results can be found. For $n=0$, i.e. $F_0=\IP^1\times \IP^1$
we note that the threepoint functions are given in the $z$ variables by
\eqn\couplings{\eqalign{
C_{111}={\Delta_2 -16 z_1 (1+z_1)\over 4 z_1^3 \Delta},\qquad
C_{112}={16 z_1^2-\Delta_2 \over 4 z_1^2 z_2 \Delta},\qquad
C_{122}={16 z_2^2-\Delta_1 \over 4 z_1 z_2^2 \Delta},}}
where $\Delta_i=(1-4 z_i)$ and
$$\Delta=1 - 8 (z_1 + z_2) +16 (z_1-z_2)^2$$
is the conifold discriminant. Other three-point function follow
for $F_0$ by symmetry. Generally these couplings of the local models
can be obtained
from the compact elliptic fibration over $F_n$ with fiber $X_6(1,2,3)$
by a limiting procedure. This compact Calabi-Yau has three complexified
volumes: $t_E$, roughly the volume of the elliptic fiber, and $t_B$ and $t_F$
the volume of base and the fiber of the Hirzebruch surface $F_n$. They
correspond to the Mori cone generators $(-6,3,2,1    ,0,0,0,0)$,
$( 0,0,0,(n-2),1,1,-n,0)$, $( 0,0,0,   -2,0,0,1,1)$ and fulfill in large
complex structure limit $\log(z_a)=t_a$. It turns out that the limit of is not given
by $t_E\rightarrow \infty$ but rather by
$\tilde t_E=(t_E-{K\cdot B \over 8} t_B-{K\cdot F \over 8}t_F)\rightarrow \infty$

With these couplings we find a particular solution to \propeq\ by choosing
\eqn\ambiguityproppIpI{\eqalign{
f^{1}_{12}&=-{1\over 4 z_2},
\qquad f^{2}_{12}=-{1\over 4 z_1}, \cr
f^{1}_{11}&=-{1\over  z_1},
\qquad f^{2}_{22}=-{1\over  z_2}, \cr}}
where the rest are either related by symmetry to the above or zero. Note that
this simple choice of the integration constants implies
algebraic relations between the Christoffel symbols in the $z$ coordinates
in the holomorphic limit
$$
\lim_{{\bar z}\rightarrow 0}
\Gamma_{z_bz_c}^{z_a}=\lim_{{\bar z}\rightarrow 0}
G^{z_a{\bar z}_e} \partial_{z_b} G_{{\bar z}_e,z_c}=
{\partial z_a\over \partial t_e} {\partial \over \partial z_b}
{\partial t_e\over \partial z_c}.
$$These relations are due the fact that only one transcendental mirror
map exists. In particular the following relation between the
mirror maps is easily shown from the Picard-Fuchs equations
\eqn\onemirrormap{{z_1\over z_2}={{q_1}\over q_2}}
with $q_i=e^{-t_i}$. With this we can obtain the general solution to
the integrability constraints as a rational relation between the
$f^{i}_{jk}$ as
$$\eqalign{
f^1_{11}&={6 z_s-1\over z_1 (1-4 z_s)}+{8 f^1_{12} z_2 z_s \over z_1 (1- 4 z_s)}-f^1_{22} {z_2^2 \over z_1^2},\cr
f^2_{11}&={z_2 (6 z_s-1)\over z_1^2 (1-4 z_s)}+{8 f^2_{12} z_2 z_s \over z_1 (1- 4 z_s)}-f^2_{22} {z_2^2 \over z_1^2},}
$$
where $z_s=z_1+z_2$.

Claim: The remaining degrees of freedom in the choice of
$f^{i}_{jk}$ can always be used to set all but one $S^{kk}$ to zero. 
This has been checked for $F_0$, see also \HosonoXJ, $F_1$ and  $F_2$ and is 
probably true more generally\foot{It would be interesting to check for compact Calabi-Yau.}.

In view of \onemirrormap\ we can see this most easily for $F_0$ in the variables
\eqn\newvariables{\eqalign{
z&={z_1\over z_2},\qquad q={q_1\over q_2},\cr
Z&=z_2,\qquad Q=q_2,}}
in which some Christoffel symbols are rational
$$\Gamma^{1}_{11}={1\over z}, \qquad  \Gamma^{1}_{12}=0, \qquad \Gamma^{1}_{22}=0$$
We can set $f^{1}_{11}=-{1\over z}$,  $f^{1}_{12}=0$ and $f^{1}_{22}=0$ so that
by
$$S^{ik}=(C_p^{-1})^{kl}(\Gamma^{i}_{pl}+f^{i}_{pl})$$
we have $S^{1p}=0$. Because of $S^{2k}=(C_p^{-1})^{kl}(\Gamma^{2}_{pl}+f^{2}_{pl})$
we must ensure that there is a rational relation between the $\Gamma^{2}_{kl}$
and rational choice of  $f^{2}_{kl}$, which is compatible with $S^{12}=S^{21}=0$.
A particular choice corresponding to \ambiguityproppIpI\ is
given by $f^1_{11}=0$, $f^2_{12}=-{1\over 4 z}$ and $f^2_{22}=-{3\over 2 Z}$.

We need the  propagator in the local orbifold  coordinates. 
It follows from the tensorial transformation law of the left 
hand side of \propeq\ and the transformation of the Christoffel on the right that a possible
choice of the ambiguity at the orbifold is given by 
${\tilde f}^{a}_{bc}={\partial x_a\over \partial z_l} 
\left(\partial z_l\over \partial x_b \partial x_c\right) + 
{\partial x_a\over \partial z_j}  
{\partial z_k\over \partial x_b}  
{\partial z_l\over \partial x_c} f^j_{kl}$, where ${ f}^i_{kl}$ are 
the ambiguities \ambiguityproppIpI\ and the transformation is given 
by \cov . This formula holds since the $s_i$ and  $t_i$ are
related by a $GL(2,\IC)$ transformation  and yields ${\tilde f}^1_{11}={1\over 1-x_1}$, 
${\tilde f}^1_{12}={\tilde f}^1_{22}={\tilde f}^2_{22}=0$, ${\tilde f}^2_{11}=-{x_2 \over 2 x_1(1-x_1)}$ and
${\tilde f}^2_{12}={4- 3 x_1\over 4 x_1( 1-x_1)}$. Note that $\Gamma^{1}_{11}=-{1\over (1-x_1)}$ 
in the $x$ coordinates and we get $S^{11}=S^{12}=S^{21}=0$. The only nonvanishing 
propagator in the $s_1,s_2$ coordinates is 
\eqn\propexp{S^{22} = {1\over 16} (s_2-s_1)(s_1+s_2) + {1\over 6144} ((s_1-s_2)(s_1+s_2)(s_1^2-5s_2^2))+ O(s^6)\ . }      
The fact that only one propagator contributes allows a consistency check or an 
alternative way of deriving the propagator. Namely by noting that 
the holomorphic anomaly equation for the genus 1 partition function
can be either derived using the contact terms in topological 
field theory or more geometrically via a generalized determinant 
calculation similar as in Quillen's work
\eqn\holanomalyfi{\partial_i \bar \partial_{\bar j} F^{(1)}=
{1\over 2} C_{ikl} {\bar C}_{\bar j \bar k \bar l} e^{2K} G^{\bar k k} G^{l\bar l}-
\left({\chi\over 24}-1\right) G_{i\bar j}.}
\ifig\fan{Degeneration of the marked torus.}
{\epsfxsize 2truein\epsfbox{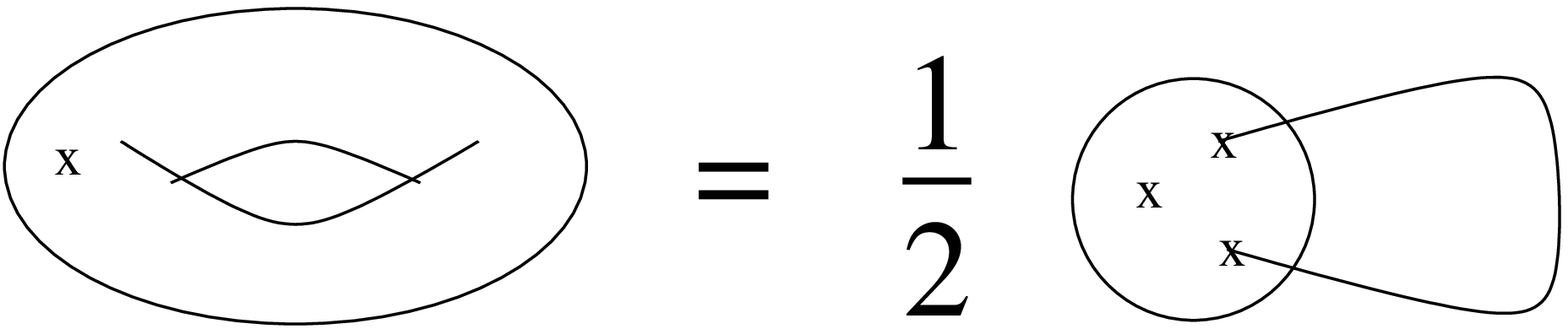}} 
If we specialize this to the non-compact case the last term becomes irrelevant in the holomorphic limit. 
Using the definition of the propagator ${\bar\partial}_{\bar j} S^{kl}:=
{\bar C}_{\bar j \bar k \bar l} e^{2K} G^{\bar k k} G^{l\bar l}$ and the fact that $C_{ikl}$ is 
truly holomorphic we may write this as 
\eqn\holfi{{\bar \partial}_{\bar j} \left[\partial_i F^1-{1\over 2} C_{ikl} S^{kl}\right]=0\ .}
This is the easiest example of the Feynman graph expansion of the anomaly equation, see \fan.

The result is that $F^{(1)}$ can be integrated in the holomorphic limit from 
\eqn\ftopi{\partial_{t_i} F^{(1)}= S^{jk} \partial_i \partial_j \partial_k F^{(0)} +\partial_{t_i} \sum_{r=1}^s a_r \log(\Delta_r)\ , }
where $\Delta_r=0$ are the various singular divisors in the moduli space and 
$\sum_{r=1}^s a_r \log( \Delta_r )$ parameterize the holomorphic ambiguity. Since 
only $S^{22}$ is nonzero we can invert the equation  \ftopi\  
and obtain \propexp\ from the knowledge of $F^{(0)},F^{(1)}$. A singular 
behavior of $S^{22}$ at the discriminant can be absorbed by choosing the 
$a_r$ appropriately. In our case the only nonzero $a_r$ will be 
$a_{con}={1\over 12}$ in order to recover 
the previous gauge choice \propexp\ \foot{In fact one can drop the ambiguity  
part $\partial_{t_i} \sum_{r=1}^s a_r \log(\Delta_r)$  in \ftopi\ altogether. 
This corresponds merely to gauge choice of the propagator which leads to a different form of 
the ambiguity at higher genus.}. We have fixed the holomorphic ambiguity up to genus three. Using 
the transformation properties of the ambiguities this allows to calculate $F^{(g)}$, $g=0,1,2,3$ 
at all points in the moduli space. We checked that the expansion at large complex structure matches the 
Gromov-Witten invariants in \amv\ and  \HosonoXJ .

\listrefs
\bye